%% file: main.tex
\documentclass[12pt]{iopart}

\RequirePackage[colorlinks=true,allcolors=black]{hyperref}
\usepackage{doi}
\usepackage{subeqn}
\usepackage{parskip}
\usepackage{graphicx}
\usepackage{subfiles}
\usepackage[ruled,vlined]{algorithm2e}

\SetKwInput{KwDesc}{Description}
\SetKwInput{KwInput}{Input}
\SetKwInput{KwOutput}{Output}
\SetKwInput{KwRequire}{Require}
\SetKwBlock{procedure}{procedure}{end\,procedure}
\newcommand{\Ids}[1]{\texttt{#1}}
\newcommand{\procedurename}[2]{\ProcNameSty{\ProcNameFnt \textsc{#1}(#2)}}
\newcommand{\Call}[2]{\ProcNameSty{\ProcNameFnt \textsc{#1}($#2$)}}

\newcommand{\State}[1]{#1 \;}
\DontPrintSemicolon
\LinesNumbered

\begin{document}

\title{Novel approach of exploring ASEP-like models through the Yang Baxter Equation}

\author{
Suvendu Barik$^{1\star}$,
Alexander. S. Garkun$^2$ and
Vladimir Gritsev$^{1}$}

\address{$^1$ Institute of Physics, University of Amsterdam, The Netherlands}
\address{$^2$ Department of Physics, Belarusian State University, Belarus}
\ead{s.k.barik@uva.nl}

\begin{abstract}We explore the algebraic structure of a particular ansatz of Yang Baxter Equation which is inspired from the Bethe Ansatz treatment of the ASEP spin-model. Various classes of Hamiltonian density arriving from two types of R-Matrices are found which also appear as solutions of constant YBE. We identify the idempotent and nilpotent categories of such constant R-Matrices and perform a rank-1 numerical search for the lowest dimension. A summary of finalised results reveals general non-hermitian spin-1/2 chain models.
 \\ \noindent{\it Keywords:\/ Yang-Baxter Integrablity, Non-Hermitian Physics}  
\end{abstract}

\submitto{\jpa}

\section{Introduction}
\subfile{section-introduction.tex}

\section{The R-Matrix and an algebra}
\subfile{section-rmatrix.tex}

\section{Numerical analysis}
\subfile{section-numcomp.tex}

\subfile{section-Xreps.tex}

\section{Conclusion}
\subfile{section-conclusion.tex}

\appendix
\subfile{section-appendices.tex}

\newpage
\section*{References}
\bibliographystyle{iopart-num}
\bibliography{refs}{}

\end{document}

%% file: section-introduction.tex
Recent decades have witnessed huge advance in our understanding of non-equilibrium classical and quantum systems, especially in one dimension. A large part of this advance is based on exact results related to integrability. One of the paradigmatic class of models in this area is a class of asymmetric simple exclusion processes (ASEP). Integrable models of these types are important in studies of integrable probability and interacting particle systems. It is an example of a solvable stochastic interface growth model, which gives rise to the Kardar-Parisi-Zhang equation \cite{Gwa92},\cite{Gwa92_2},\cite{Kim}; see surveys by \cite{Golinelli_2006}. Other integrable models with similar properties include the stochastic six vertex model. The particular case of the open ASEP is defined as the following interacting particle system. Particles occupy sites in a finite chain $\{1, \dots , N\}$ for some $N$, and they jump left at rate $q$ and right at rate $p$. Moreover, particles are inserted into site 1 at rate $\alpha$ and removed from there at rate $\gamma$, while at site $N$ insertion occurs at rate $\delta$ and removal at rate $\beta$. All moves that violate the rule of at most one particle per site at a given time are excluded. These models have found various applications. Exactly solvable cases that were found in 90's \cite{Gwa92} were generalized and extended further by many authors  \cite{ALCARAZ1994250}, \cite{Fabian}, \cite{DERRIDA199865}, \cite{Schtz20011E}, \cite{deGier_2006}. These models can be mapped to non-hermitian spin chains and in many cases have hidden algebraic structures, like e.g. Temperley-Lieb, Hecke \cite{ALCARAZ1994250}, q-deformed or more general quadratic algebras \cite{Fabian}.

In this paper we extend the class of solvable ASEP models related to certain algebraic structures (specified below) and corresponding spin chains. We obtain solutions of the Yang-Baxter equation which correspond to non-full rank matrices, generalizing results obtained in \cite{garkun2024new}.

%% file: section-rmatrix.tex
\subsection{Yang Baxter Equation}
\label{sec:Rsymm}
The focus in this paper is to study the Yang Baxter Equation (YBE)
\begin{equation}
\label{eq:RRR}
\eqalign{\mathcal{R}_{12}\left(f(u_1,u_2)\right)\mathcal{R}_{13}\left(f(u_1,u_3)\right)\mathcal{R}_{23}\left(f(u_2,u_3)\right) = \cr\mathcal{R}_{23}\left(f(u_2,u_3)\right)\mathcal{R}_{13}\left(f(u_1,u_3)\right)\mathcal{R}_{12}\left(f(u_1,u_2)\right)}
\end{equation}
where we parameterise the rapidities with a general function $f(x,y)$. To look into R-Matrices similar to that of the ASEP model, we consider the following R-Matrix ansatz
\begin{equation}
\label{eq:Rmatr}
\mathcal{R}\left(f\left(x,y\right)\right)=\mathcal{P}\left(I+f\left(x,y\right)\mathcal{M}\right)
\end{equation}
where $\mathcal{M}$ is a square matrix with complex coefficients and $\mathcal{P}$ is the transposition operator. By imposing $\lim_{y\rightarrow x}f(x,y)=0$, $\mathcal{R}$ then satisfies the regularity condition. 

We construct the below transfer matrix of $N$ lattice sites
\begin{equation}
\label{eq:Transfermatrix}
\tau(x,y) = \Tr_{\mathcal{A}}(\mathcal{R}_{0,1}(f(x,y))\mathcal{R}_{0,2}(f(x,y))\cdots \mathcal{R}_{0,N}(f(x,y)))
\end{equation}
where each $\mathcal{R}_{0,n}$ acts on $\mathcal{A}\otimes\mathcal{H}_{n}$, with $\mathcal{H}_{n}$ being the Hilbert space for the local site $n$. $\mathcal{A}$ is an auxiliary vector space isomorphic to $\mathcal{H}_{n}$. By considering the following ordering of limits $ : (y\rightarrow x, x \rightarrow 0)$, we calculate the first integral of motion 
\begin{equation}
\eqalign{
T = \tau(0,0) &= \lim_{x \rightarrow 0}\lim_{y\rightarrow x}\Tr_{\mathcal{A}}(\mathcal{R}(f(x,y))_{0,1},\dots,\mathcal{R}(f(x,y))_{0,N}) \cr
&=\Tr_{\mathcal{A}}(\mathcal{P}_{0,1},\dots,\mathcal{P}_{0,N}) \cr
&=\Tr_{\mathcal{A}}(\mathcal{P}_{1,0})\mathcal{P}_{1,2},\dots,\mathcal{P}_{1,N} \cr
&=\mathcal{P}_{1,2},\dots,\mathcal{P}_{1,N}
}   
\end{equation}
which is the translation operator satisfying $T^{N}=\mathbf{I}$. It generates translations in a periodic lattice. Considering the first derivative with respect to $x$ on $\tau$ reveals the second integral of motion 
\begin{equation}
H = \lim_{x \rightarrow 0}\lim_{y\rightarrow x}\sum_{k=1}^{N}\mathcal{P}_{k,k+1}\left(\frac{d\left(\mathcal{R}_{k,k+1}(f\left(x,y\right))\right)}{dx}\right),
\end{equation}

which is the nearest-neighbouring Hamiltonian. By identifying $df(x,y)/dx$ with constant $\alpha$ after taking the limits, the Hamiltonian simplifies as 
\begin{equation}
H = \alpha \sum_{k=1}^{N} \mathcal{M}_{k,k+1}.
\end{equation}
One of the key observation is that $\mathcal{M}$ represents the Hamiltonian density. Expanding \eref{eq:RRR} by substituting \eref{eq:Rmatr} and with further simplifications gives the constraint on $\mathcal{M}$
\begin{equation}
\label{eq:Mconstr}
\eqalign{
&\left(f_{12}+f_{23}-f_{13}\right)\left(\mathcal{M}_{23}-\mathcal{M}_{12}\right)+f_{12}f_{23}\left(\mathcal{M}_{23}^2-\mathcal{M}_{12}^2\right)\cr+&f_{12}f_{13}f_{23}\left(\mathcal{M}_{23}\mathcal{M}_{12}\mathcal{M}_{23}-\mathcal{M}_{12}\mathcal{M}_{23}\mathcal{M}_{12}\right)=0
}    
\end{equation}
where $f_{ij}\equiv f\left(u_i,u_j\right)$. 

\subsection{Algebra of the Hamiltonian density}
Assuming that the parameterisation $f(x,y)$ does not diverge as $y \rightarrow x$ and satisfies $\lim_{y\rightarrow x}f(x,y) = 0$, we first consider taking possible pairs of the spectral parameters $u_1, u_2, u_3$ casewise in \eref{eq:Mconstr}. For $u_2 \rightarrow u_1$ and $u_3\rightarrow u_2$ the constraint vanishes. By taking $u_3\rightarrow u_1$ we reveal a non-trivial condition which is given by
\begin{equation}
\label{eq:firstconstr}
\left(\mathcal{M}_{23}^2-\mathcal{M}_{12}^2\right)=\frac{\left(f_{12}+f_{21}\right)}{f_{12}f_{21}}\left(\mathcal{M}_{12}-\mathcal{M}_{23}\right).
\end{equation}
Since this expression has to be true for all values of $u_{1},u_{2}$, we impose
\begin{equation}
\label{eq:fconst1}
\frac{\left(f_{12}+f_{21}\right)}{f_{12}f_{21}}=\omega,\;\;\omega\in\mathbf{C}.
\end{equation}
Using \eref{eq:firstconstr} in \eref{eq:Mconstr} and rearranging the expression 
\begin{equation}
\fl \left(\mathcal{M}_{23}\mathcal{M}_{12}\mathcal{M}_{23}-\mathcal{M}_{12}\mathcal{M}_{23}\mathcal{M}_{12}\right)=\frac{\left(f_{12}+f_{23}-f_{13}-\omega f_{12}f_{23}\right)}{f_{12}f_{13}f_{23}}\left(\mathcal{M}_{23}-\mathcal{M}_{12}\right)
\end{equation}
we have the second constraint that
\begin{equation}
\label{eq:fconst2}
\frac{1}{f_{12}f_{13}f_{23}}\left(f_{12}+f_{23}-f_{13}-\omega f_{12}f_{23}\right)=\kappa,\;\;\kappa\in\mathbf{C}.
\end{equation}

\noindent In this paper we will consider the following parameterisation
\begin{equation}
f(x,y)=\frac{x-y}{\sum_{i,j=0}^{N}d_{ij}x^{i}y^{j}}    
\end{equation}
for arbitrary $d_{ij}$, which automatically satisfies $\lim_{y\rightarrow x}f(x,y)=0$. After using \eref{eq:fconst1} and \eref{eq:fconst2}, we find
\begin{equation}
\label{eq:RRRpar}
f(x,y)=\frac{x-y}{c_0^{2}+c_{0}c_{1}(x+y)+c_{1}^{2}xy+\omega x+\left(\frac{c_1}{c_0}\omega-\frac{\kappa}{c_{0}^{2}}\right)xy}
\end{equation}
where $c_{0}, c_{1}$ are free complex constants. The calculations leading towards the expression is given in \ref{appendix:A}. 

We arrive at two constraints on $\mathcal{M}$, which are
\begin{equation}
\label{eq:constr}
\eqalign{
&\left(\mathcal{M}_{23}^2-\mathcal{M}_{12}^2\right)=\omega\left(\mathcal{M}_{12}-\mathcal{M}_{23}\right), \cr
&\left(\mathcal{M}_{23}\mathcal{M}_{12}\mathcal{M}_{23}-\mathcal{M}_{12}\mathcal{M}_{23}\mathcal{M}_{12}\right)=\kappa\left(\mathcal{M}_{12}-\mathcal{M}_{23}\right).
}
\end{equation}

\subsubsection{Generalising the constraints on \texorpdfstring{$\mathcal{M}$}{M}}

Extending \eref{eq:constr} for arbitrary site indices $(i, i+1, i+2)$ from $(1, 2, 3)$ correspondingly and using $e_{i} \equiv \mathcal{M}_{i,i+1}$, we get
\begin{equation}
\label{eq:genalgmin}
\eqalign{
e_{i}^{2}+\omega e_{i} = e_{i+1}^{2} + \omega e_{i+1}, \cr
e_{i}e_{i+1}e_{i}+\kappa e_{i} = e_{i+1}e_{i}e_{i+1}+\kappa e_{i+1}.
}
\end{equation}
The Hamiltonian $H$ becomes $\alpha\sum_{i=1}^{N} e_{i}$ with $\alpha=c_{0}^{-2}$. Looking at the first constraint, we note that
\begin{equation}
e_{1}^{2}+\omega e_{1} = e_{2}^{2} + \omega e_{2}\dots = e_{N}^{2} + \omega e_{N}
\end{equation}
which is satisfied if $ e_{i}^{2}+\omega e_{i} = \lambda\mathbf{I}$ for some complex constant $\lambda$. Similarly, we impose
\begin{equation}
\label{eq:intertwining}
e_{i}e_{i+1}e_{i}+\kappa e_{i} = e_{i+1}e_{i}e_{i+1}+\kappa e_{i+1} \equiv t_{i,i+1}    
\end{equation}
where we define $t_{i,i+1}$ as a three site operator acting on $\mathcal{H}_{i}\otimes\mathcal{H}_{i+1}\otimes\mathcal{H}_{i+2}$ which is invariant under $i \leftrightarrow i+1$ exchange. 

In the end, the algebraic conditions that $e_{i}$ needs to satisfy, starting from \eref{eq:Mconstr} are
\begin{equation}
\label{eq:alg}
\eqalign{
e_{i}^{2} = \lambda\mathbf{I} - \omega e_{i}, \cr
e_{i}e_{i+1}e_{i} = t_{i,i+1} - \kappa e_{i}, \cr
e_{i+1}e_{i}e_{i+1} = t_{i,i+1} - \kappa e_{i+1}.
}   
\end{equation}
For different conditions on  $\kappa, \lambda$ and $t_{i,i+1}$, the above condition pinpoint to various algebraic structures that the generators are required to satisfy.

\subsection{Exploring the algebraic conditions}
We mention the important aspects of each conditions in \eref{eq:genalgmin}. The first condition
\begin{equation}
e_{i}^{2}+\omega e_{i}-\lambda\mathbf{I} = 0
\end{equation}
is also known as the eigenvalue problem. One can write it in the following factorised form - 
\begin{equation}
\label{eq:factore}
(e_{i}-\nu_{+}\mathbf{I})(e_{i}-\nu_{-}\mathbf{I})=\mathbf{0},\;\;\nu_{\pm}=\frac{1}{2}(-\omega\pm c_{\omega}(\lambda)),
\end{equation}
where we will use $c_{\omega}(x)\equiv\sqrt{\omega^{2}+4x}$ as a shorthand. The second condition in \eref{eq:genalgmin} is the intertwining equation, which holds the important constraints arising from the Yang Baxter Equation. By rewriting the generator $e_{i}$ as $q_{i} + \beta\mathbf{I}$  with $\beta\in\mathbf{C}$, it is rewritten into the braid equation as
\begin{equation}
\label{eq:braidequiv}
q_{i}q_{i+1}q_{i}=q_{i+1}q_{i}q_{i+1},
\end{equation}
after fixing $\beta$ such that $\beta^{2}+\beta\omega-\kappa = 0$. After choosing the positive branch of the quadratic root, the eigenvalue problem \eref{eq:factore} modifies for $q_{i}$ as 
\begin{equation}
\label{eq:eigvalueq}
\left(q_{i}+\frac{1}{2}\left(c_{\omega}(\kappa)+c_{\omega}(\lambda)\right)\mathbf{I}\right)\left(q_{i}+\frac{1}{2}\left(c_{\omega}(\kappa)-c_{\omega}(\lambda)\right)\mathbf{I}\right) =0.  
\end{equation}

\subsubsection{Solution classes}In order to identify classes of \eref{eq:eigvalueq}, we consider the constraints on $c_{\omega}(\kappa)$ and $c_{\omega}(\lambda)$. For the case when $c_{\omega}(\kappa) \neq c_{\omega}(\lambda)$, we arrive to
\begin{equation}
(\tilde{q}_{i}+\mathbf{I})(\tilde{q}_{i}-\theta\mathbf{I}) = 0,\;\;\tilde{q}_{i} = \frac{2q_{i}}{(c_{\omega}(\kappa)-c_{\omega}(\lambda))},
\end{equation}
with
\begin{equation}
-\theta = \frac{c_{\omega}(\kappa)+c_{\omega}(\lambda)}{c_{\omega}(\kappa)-c_{\omega}(\lambda)},
\end{equation}

\noindent which represents the familiar Iwahori-Hecke algebra \cite{iwahori1964},\cite{jones1987}, with $\theta\neq 0$, while writing \eref{eq:braidequiv} with some non-zero $C$
\begin{equation}
\label{eq:braidequtilde}
\tilde{q}_{i}\tilde{q}_{i+1}\tilde{q}_{i} = \tilde{q}_{i+1}\tilde{q}_{i}\tilde{q}_{i+1},\;\tilde{q}_{i} = C q_{i}.
\end{equation}
\noindent For the case of $c_{\omega}(\kappa)=c_{\omega}(\lambda)$, with $\kappa=\lambda\neq-\omega^{2}/4$ we get $\tilde{q}_{i}^{2}=\tilde{q}_{i},\;\tilde{q}_{i} = -q_{i}/c_{\omega}(\lambda),\;$ which corresponds to the idempotent generators of the braid equation. Finally for $\kappa=\lambda=-\omega^2/4$, we have $\tilde{q}_{i}^{2} = 0,\;\tilde{q}_{i}=q_{i},\;$ which represents nilpotent generators of degree 2.

The re-scaled braid equation in \eref{eq:braidequtilde} is equivalent to the constant Yang Baxter Equation (cYBE) where the constant R-Matrix $Q_{i,i+1}$ and  $\tilde{q}_{i}$ are related by $\tilde{q}_{i} = \mathcal{P}_{i,i+1}Q_{i,i+1}$. The Hamiltonian then becomes as
\begin{equation}
\label{eq:hamqtilde}
H = \frac{\alpha}{C}\sum_{i=1}^{N}\tilde{q}_{i} + N\alpha\beta\mathbf{I},\;\;\alpha=c_{0}^{-2},
\end{equation}
and the R-Matrix \eref{eq:Rmatr} as
\begin{equation}
\label{eq:Rmatrqtilde}
R_{ij}(f(x,y)) = (1+\beta f(x,y))\mathcal{P}_{ij} + \frac{f(x,y)}{C}Q_{ij}.
\end{equation}
A summary the different eigenvalue problems with the corresponding forms of $C$ and $\beta$ in \Tref{table:soltypes}. In this manner, we have transformed the problem into solving for $\tilde{q}_{i}$ satisfying \eref{eq:braidequtilde} with any of the three eigenvalue problems depending on what constraints $\lambda,\kappa$ and $\omega$ satisfy.

\begin{table}
\centering
\caption{\label{table:soltypes} Forms of possible Hamiltonian density. The $\mathbf{R}$ column \\ refers the relations which satisfies $\mathbf{R}=\mathbf{0}$.}
\begin{tabular}{@{}ccccc@{}}
\br
Type & Case & $C$ & $\beta$ & $\mathbf{R}$ \\ \mr
$A$&$\kappa\neq\lambda$ &
$2(c_{\omega}(\kappa) - c_{\omega}(\lambda))^{-1}$ &
$\frac{1}{2}(-\omega+c_{\omega}(\kappa))$ &
$(\tilde{q}_{i}+\mathbf{I})\left(\tilde{q}_{i}-\theta\mathbf{I}\right)$ \\ 
$B$&$\kappa = \lambda \neq -\frac{1}{4}\omega^{2}$ &
$-c_{\omega}(\lambda)^{-1}$ &
$\frac{1}{2}(-\omega+c_{\omega}(\lambda))$ &
$\tilde{q}^{2}-\tilde{q}$ \\ 
$C$&$\kappa = \lambda = -\frac{1}{4}\omega^{2}$ &
1 &
$-\frac{1}{2}\omega$ &
$\tilde{q}^{2}$ \\ \br
\end{tabular}
\end{table}

\subsubsection{Representation of the Hamiltonian density}
We will focus on \eref{eq:factore} with $e_{i}$ as a $N^2\times N^2$ matrix for $\dim(\mathcal{A})=N$ and provide the necessary matrix representation. By identifying $p(x)=(x-\nu_{+})(x-\nu_{-})$ as the minimal polynomial of $e_{i}$, we use the Primary Decomposition Theorem \cite{hoffman1971linear} to get
\begin{equation}
\ker(e_{i}-\nu_{+}\mathbf{I})\oplus\ker(e_{i}-\nu_{-}\mathbf{I})=\mathbf{C}^{N^2}.
\end{equation}
After applying $\dim(A\oplus B)=\dim(A)+\dim(B)$ and rank-nullity theorem \cite{hoffman1971linear},\cite{nakahara2003geometry}, we find
\begin{equation}
R_k(e_{i}-\nu_{+}\mathbf{I})+R_k(e_{i}-\nu_{-}\mathbf{I})=N^2,
\end{equation}
where $R_k(M)$ is the rank of the square matrix $M$. If we identify $\;\Lambda_{i}=e_{i}-\nu_{+}\mathbf{I}\;$ as a rank $r$ matrix, then  $\;e_{i}-\nu_{-}\mathbf{I}=\Lambda_{i}+c_{\omega}(\lambda)\mathbf{I}\;$ is a rank $N^2-r$ matrix. Then we rewrite \eref{eq:factore} as
\begin{equation}
\label{eq:Lambdaform}
\Lambda_{i}(\Lambda_{i}+c_{\omega}(\lambda)\mathbf{I})=\mathbf{0}
\end{equation}
for constructing the matrix representations of $e_{i} = \Lambda_{i} + \nu_{+}\mathbf{I}$. The essential property to notice is that for $\omega^{2}=-4\lambda$, we have $\Lambda_{i}$ as a nilpotent matrix of degree $2$. For the case where $\omega^{2}\neq-4\lambda$, we then rewrite \eref{eq:Lambdaform} with $\Phi_{i}=-c_{\omega}(\lambda)^{-1}\Lambda_{i}$ as $\Phi_{i}^{2}=\Phi_{i}$ which reveals the idempotent nature of $\Lambda_{i}$. In this fashion, we only require nilpotent and idempotent matrices to construct two possible solutions of $e_i$.

\subsubsection{General form of \texorpdfstring{$e_{i}$}{ei} and intertwining relations}

With $D=N^2$, we have the desired forms of $e_{i}$ as below 
\begin{equation}
\label{eq:eiforms}
e_{i} = \cases{\mathbf{N}_{r} - \frac{\omega}{2}\mathbf{I} & $\omega^{2}=-4\lambda$, \\ -c_{\omega}(\lambda)\mathbf{B}_{r} + \nu_{+}\mathbf{I} & $\omega^{2}\neq-4\lambda$,}
\end{equation}

\noindent with $\mathbf{N}_{r}$ and $\mathbf{B}_{r}$ acting on $\mathcal{H}_{i}\otimes\mathcal{H}_{i+1}$ are the order-2 Nilpotent and Idempotent matrices of rank $r$ respectively. By using \eref{eq:intertwining}, we find the intertwining constraints that both of them are required to satisfy as
\begin{equation}
\mathbf{Z}\otimes\mathbf{I}\cdot\mathbf{I}\otimes \mathbf{Z}\cdot\mathbf{Z}\otimes\mathbf{I} - \mathbf{I}\otimes \mathbf{Z}\cdot\mathbf{Z}\otimes\mathbf{I}\cdot\mathbf{I}\otimes \mathbf{Z} = f_{\lambda,\kappa}[\mathbf{I}\otimes \mathbf{Z} - \mathbf{Z}\otimes\mathbf{I}],
\end{equation}

where
\begin{equation}
\mathbf{Z} = \cases{\mathbf{N}_{r} & $f_{\lambda,\kappa}=\frac{1}{4}(\omega^2+4\kappa)$, \\ \mathbf{B}_{r} & $f_{\lambda,\kappa}= (\kappa-\lambda)(\omega^{2}+4\lambda)^{-1}$.}
\end{equation}

In the end we categorically write all possible forms of $\tilde{q}_{i}$ using Table \ref{table:soltypes} as
\begin{equation}
\label{eq:soltypes}
\tilde{q}_{i} = \cases{
\mathbf{N}_{r} & $\omega^{2}=-4\lambda=-4\kappa$ (BN), \\
\mathbf{B}_{r} & $\omega^{2}\neq-4\lambda=-4\kappa$ (BI), \\
\frac{2}{c_{\omega}(\kappa)}\mathbf{N}_{r}-\mathbf{I} & $\omega^{2}=-4\lambda\neq-4\kappa$ (HN), \\
\frac{2c_{\omega}(\lambda)}{c_{\omega}(\lambda)-c_{\omega}(\kappa)}\mathbf{B}_{r}-\mathbf{I} & $\omega^{2}\neq-4\lambda\neq-4\kappa$ (HI);\\
}\end{equation}

\noindent which we will use to solve for $\mathbf{N}_{r}$ and $\mathbf{B}_{r}$ through the braid equation \eref{eq:braidequtilde} and identify the Hamiltonian \eref{eq:hamqtilde} and the R-Matrix \eref{eq:Rmatrqtilde}. Any solution of type BI and BN from \eref{eq:soltypes} are low-rank matrices. For an idempotent matrix, the rank $r$ is less than the dimension $N$. If $r=N$ then it corresponds to the identity matrix. For nilpotent matrix (of order 2), we have $r\leq N/2$. Hence we have shown two classes of low-rank solutions of cYBE related physically through \eref{eq:Rmatrqtilde}. Solutions from HI and HN classes may not necessarily have $r<N$.

\subsection{The three site operator}

From \eref{eq:alg}, the three site operator $t_{i,i+1}$ can be rewritten into the following symmetrised form with respect to $q_{i}$ as
\begin{equation}
\label{eq:tsop}
t_{i,i+1} = q_{i}q_{i+1}q_{i} + \beta\{q_{i},q_{i+1}\} +\beta^{2}(q_{i}+q_{i+1}) + \beta(\beta^{2}+\lambda)\mathbf{I}
\end{equation}
after setting the value of $\beta$ which is considered in \eref{eq:eigvalueq}. A multiplication of the three site operator with $q_{i}-q_{i+1}$ then simplifies towards the following property
\begin{equation}
t_{i,i+1}(q_{i}-q_{i+1}) = \lambda[q_{i},q_{i+1}]
\end{equation}
and similarly 
\begin{equation}
t_{i,i+1}(q_{i}+q_{i+1}+2(\beta+\omega)\mathbf{I}) = \lambda\left(2\kappa\mathbf{I}+\{q_{i}+\beta\mathbf{I},\;q_{i+1}+\beta\mathbf{I}\}\right). 
\end{equation}

\subsubsection{Deducing the Temperley Lieb Algebra}
Now we can use the properties of the operator to show that if $t_{i,i+1}=\mathbf{0}$, then $\lambda=0$ for non-trivial $q_{i}$.Let $t_{i,i+1}=\mathbf{0}$. If $\lambda\neq 0$, then 
\begin{equation}
[q_{i},q_{i+1}] = 0,\,\,\{e_{i},e_{i+1}\} = -2\kappa\mathbf{I},\,\,\forall i.  
\end{equation}

From the condition of the commutator, $q_{i}$ needs to be a single lattice-site term. From the anti-commutator relation, we then write
\begin{equation}
q_{i}q_{i+1} + \beta(q_{i}+q_{i+1}) + (\kappa+\beta^{2})\mathbf{I} = 0,\,\forall i
\end{equation}

which is only possible if each $q_i$ is proportional to the identity $\mathbf{I}$. Hence for the non-trivial generator nulling out the three site operator, we need $\lambda=0$.

Then it follows that a non-trivial representation of $e_{i}$ satisfies the Temperley Lieb algebra if $t_{i,i+1}=\mathbf{0}$ as one can check from \eref{eq:alg}. We have used it for simplifying the numerical computations in the next sections where we focus on providing a list of lowest dimensional solutions ($N=2$) of the R-Matrix \eref{eq:Rmatr}.

%% file: section-numcomp.tex
We have computed the relevant rank-1 Idempotent and degree-2 Nilpotent square matrices of dimension $N^{2}=4$ for the possible forms of $e_{i}$ through the classification given in \eref{eq:soltypes}, which is enough to reconstruct the R-Matrix \eref{eq:Rmatrqtilde}. The choice of working with rank-1 models lies in the idea of avoiding the sets of equations which are computationally difficult to solve. We also provide how to construct these matrices numerically in \ref{appendix:B}.

Another reason is that $\mathbf{A}_{r}$ from \eref{eq:BNform} represents a sum of various rank-1 matrices which are linearly independent. Considering a rank $r>1$ involves $2N^{2}r$  variables with $r^{2}$ additional constraints from \eref{eq:idcond} or \eref{eq:nlcond}. One may think of rank-$m$ models as an added generalisation to a rank-$n$ case for $n<m$. Hence solving for the lowest rank is a key step.

In this section, we will describe the computational workflow involved in the analysis and the methods for simplifying the results.

\subsection{Computational workflow}
Our numerical methodology is divided into three phases.
\subsubsection*{Main Computation}
In the specific case of $N=2$, we are able to directly compute the solutions of \eref{eq:soltypes} from the braid equation \eref{eq:braidequtilde} with $r=1$. Fortunately the use of Gröbner basis for decomposing a maximal $N^{6}+r^{2}=65$ over-deterministic equations for $2N^{2}r=8$ unknowns are tenable with Mathematica packages. Hence for every solution pool (BI, BN, HI, HN), we were able to gather results by using \verb|Solve[]| and \verb|Reduce[]| modules available in the package.
\subsubsection*{Removing redundant results}
The next step is to remove redundant solutions from the gathered results. Using the symmetries of the R-Matrix mentioned in \ref{appendix:C}, we have made routines to identify repeating solutions. The pseudocodes of the programs used are given in \ref{appendix:D}. 

For the proceeding step, we introduce the structure matrix
\begin{equation}
S(M) = [s_{ij}],\;\;s_{ij} = \cases{1 & if $m_{ij}\neq 0$\\0 & if $m_{ij}= 0$}
\end{equation}
for identifying the non-zero elements of the matrix $M$. We call any solution whose structure matrix have no zero elements as \textit{full-case} matrices. 

We decided to break full-case solutions into various subcases having some zero elements. For this, we made valid substitution of their free constant parameters to zero. All of them are then gathered with the rest of the results and were checked together for repetitions.

The symmetries of the R-Matrix act equivalently on the Nilpotent/Idempotent matrices due to its form in \eref{eq:Rmatrqtilde}. Hence we can remove repetitions within each solution categories in \eref{eq:soltypes} while ensuring the distinction among them.

\subsubsection*{Post Simplifications}
At the final step, we nullified the three site operator in \eref{eq:tsop} for every solution in order to simplify our results, which then fulfil the Temperley-Lieb Algebra after writing them in the form of $e_{i}$. External parameters like $\lambda, \omega,\kappa$ are reduced case-wise if they do not contribute in the solution. 

\subsection{Results of rank-1 models}
Here we present all the rank-1 models which have some zero matrix terms. For simplicity, we also mention the subcases of \eref{eq:alg} that each result classes fulfil. We name each classes accordingly to which of eigenvalue problem and intertwining relation they satisfy. The parameters $c_1,c_2$ and $c_3$ are free in these solutions. 

\subsubsection{Braid-Nilpotent}
The solutions is of form $e_{i} = \mathbf{N}_{1}$, satisfying 
\begin{equation} 
e_{i}^{2} = \mathbf{0},\;\; e_{i}e_{i+1}e_{i} = \mathbf{0},\;\; e_{i+1}e_{i}e_{i+1} = \mathbf{0}.
\end{equation}
The list of these $\mathbf{N}_{1}$ are in \eref{eq:BN}.

\begin{equation}
\label{eq:BN}
\fl \eqalign{
M_{BN}(a) =  \left(\begin{array}{cccc}
0 & -c_{2} & 0 & -c_{2}^{2}c_{1}^{-1} \\
0 & 0 & 0 & 0 \\
0 & c_{1} & 0 & c_{2} \\
0 & 0 & 0 & 0
\end{array}\right)\;\;
M_{BN}(b) = \left(\begin{array}{cccc}
0 & 0 & 0 & c_{3} \\
0 & 0 & 0 & c_{2} \\
0 & 0 & 0 & c_{1} \\
0 & 0 & 0 & 0
\end{array}\right) 
}    
\end{equation}

\subsubsection{Braid-Idempotent}
The solutions is of form $e_{i} = \mathbf{B}_{1}$, satisfying 
\begin{equation}
e_{i}^{2} = e_{i},\;\; e_{i}e_{i+1}e_{i} = \mathbf{0},\;\; e_{i+1}e_{i}e_{i+1} = \mathbf{0}.
\end{equation}
The list of these $\mathbf{B}_{1}$ are in \eref{eq:BI}.

\begin{equation}
\label{eq:BI}
\fl \eqalign{M_{BI}(a) = \left(\begin{array}{cccc}
0 & 0 & 0 & 0 \\
-c_{2} & 1-c_{1} & -c_{1} & c_{1}(1-c_{1})c_{2}^{-1} \\
c_{2} & -1+c_{1} & c_{1} & c_{1}(c_{1}-1)c_{2}^{-1} \\
0 & 0 & 0 & 0\end{array}\right)\cr
M_{BI}(b) =  \left(\begin{array}{cccc}
0 & c_{2} & 0 & -c_{2}^{2}(c_{1}+1)^{-1} \\
0 & 1 & 0 & -c_{2}(c_{1}+1)^{-1} \\
0 & c_{1} & 0 & -c_{1}c_{2}(c_{1}+1)^{-1} \\
0 & 0 & 0 & 0
\end{array}\right)
M_{BI}(c) = \left(\begin{array}{cccc}
0 & 0 & c_{2}c_{1}^{-1} & c_{2} \\
0 & 0 & 0 & 0 \\
0 & 0 & 1 & c_{1} \\
0 & 0 & 0 & 0
\end{array}\right)\cr M_{BI}(d) = \left(\begin{array}{cccc}
0 & 0 & -c_{2} & c_{2}c_{1}^{-1} \\
0 & 0 & -(c_{1}c_{2}+1) & (c_{1}c_{2}+1)c_{1}^{-1} \\
0 & 0 & -c_{1}c_{2} & c_{2} \\
0 & 0 & -c_{1}(c_{1}c_{2}+1) & c_{1}c_{2}+1
\end{array}\right)  
}    
\end{equation}

There is a model of form $e_{i} = \mathbf{B}_{1}$ 
\begin{equation}
\fl M_{BI}(e) = \left(\begin{array}{cccc}
0 & 0 & 0 & c_{1}^{2} \\
0 & 0 & 0 & c_1 \\
0 & 0 & 0 & c_1 \\
0 & 0 & 0 & 1
\end{array}\right) 
\end{equation}
which satisfies $e_{i}^{2} = e_{i},\;\; e_{i}e_{i+1}e_{i} = e_{i+1}e_{i}e_{i+1} \neq \mathbf{0}$.

\subsubsection{Hecke-Nilpotent}
The solutions is of form $e_{i} = \mathbf{N}_{1}$, satisfying 
\begin{equation}
e_{i}^{2} = 0,\;\; e_{i}e_{i+1}e_{i} = -\kappa e_{i},\;\; e_{i+1}e_{i}e_{i+1} = -\kappa e_{i+1}.
\end{equation}
The list of these $\mathbf{N}_{1}$ are in \eref{eq:HN}.

\begin{equation}
\label{eq:HN}
\fl\eqalign{M_{HN}(a) = \left(\begin{array}{cccc}
0 & 0 & 0 & 0 \\
-c_{2} & -c_1 & -c_1 & (\kappa-c_1^2)c_2^{-1} \\
c_2 & c_1 & c_1 & (-\kappa+c_1^2)c_2^{-1} \\
0 & 0 & 0 & 0\end{array}\right)\cr
M_{HN}(b) =  \left(\begin{array}{cccc}
0 & (c_{2} \kappa^{1/2})c_{1}^{-1} & c_{2} & (q_{\kappa}c_{2}^2)c_{1}^{-1}\\
0 & \kappa^{1/2} & c_{1} & c_{2}q_{\kappa}\\
0 & -\kappa c_{1}^{-1} & -\kappa^{1/2} & -(\kappa^{1/2}q_{\kappa} c_{2})c_{1}^{-1} \\
0 & 0 & 0 & 0 \\
\end{array}\right)\cr
M_{HN}(c) = \left(\begin{array}{cccc}
-2 c_{1} & 2 c_{1}^2 c_{2}(\kappa -c_{1}^2)^{-1} & 2 c_{1}^2 c_{2}(\kappa -c_{1}^2)^{-1} & 2 c_{1} c_{2}^2(\kappa -c_{1}^2)^{-1} \\
(c_{1}^2-\kappa )c_{2}^{-1} & c_{1} & c_{1} & c_{2} \\
(c_{1}^2-\kappa )c_{2}^{-1} & c_{1} & c_{1} & c_{2} \\
0 & 0 & 0 & 0 \\
\end{array}\right)\cr
}    
\end{equation}
where $q_{\kappa}=\left(c_{1}+\kappa^{1/2}\right)\left(\kappa^{1/2}-c_{1}\right)^{-1}$.

\subsubsection{Hecke-Idempotent}
We found two types of solutions for this type. The first type is of form $e_{i}=\mathbf{B}_{1}$, satisfying
\begin{equation}
e_{i}^{2} = e_{i},\;\; e_{i}e_{i+1}e_{i} = \frac{1}{4}e_{i},\;\; e_{i+1}e_{i}e_{i+1} = \frac{1}{4} e_{i+1}.
\end{equation}
The list of these $\mathbf{B}_{1}$ are in \eref{eq:HIa}
\begin{equation}
\label{eq:HIa}
\fl \eqalign{M_{HIa}(a) = \left(\begin{array}{cccc}
0 & 0 & 0 & 0 \\
(1-2 c_{2})^2(4 c_{1})^{-1} & c_{2} & c_{2}-1 & c_{1} \\
-(1-2 c_{2})^2(4 c_{1})^{-1} & -c_{2} & 1-c_{2} & -c_{1} \\
0 & 0 & 0 & 0 \\ \end{array}\right)\cr
M_{HIa}(b) =  \frac{1}{2}\left(\begin{array}{cccc}
1& 0 & -c_{1} & c_{2} \\
c_{1}c_{2}^{-1} & 0 & -c_{1}^2c_{2}^{-1} & c_{1} \\
0 & 0 & 0 & 0 \\
c_{2}^{-1} & 0 & -c_{1}c_{2}^{-1} & 1 \\
\end{array}\right)\cr
M_{HIa}(c) = \frac{1}{2(c_{1}-c_{2})}\left(\begin{array}{cccc}
c_{1}+c_{2} & 0 & -4 & 4(c_{1}+c_{2})^{-1} \\
c_{1} (c_{1}+c_{2}) & 0 & -4 c_{1} & 4 c_{1}(c_{1}+c_{2})^{-1} \\
c_{2} (c_{1}+c_{2}) & 0 & -4 c_{2} & 4 c_{2}(c_{1}+c_{2})^{-1} \\
\frac{1}{4} (c_{1}+c_{2})^3 & 0 & -(c_{1}+c_{2})^2 & c_{1}+c_{2} \\
\end{array}\right) 
}    
\end{equation}
\vspace{0.4cm}
The second type is of form $e_{i}=\mathbf{B}_{1}$, satisfying
\begin{equation}
e_{i}^{2} = e_{i},\;\;e_{i}e_{i+1}e_{i} = -\kappa e_{i},\;\;
e_{i+1}e_{i}e_{i+1} = -\kappa e_{i+1}.
\end{equation}
The list of these $\mathbf{B}_{1}$ are in \eref{eq:HIb}
\begin{equation}
\label{eq:HIb}
\fl \eqalign{M_{HIb}(a) = \left(\begin{array}{cccc}
0 & 0 & 0 & 0 \\
-(\kappa-c_{2}^2+c_{2})c_{1}^{-1} & c_{2} & c_{2}-1 & c_{1} \\
(\kappa-c_{2}^2+c_{2})c_{1}^{-1} & -c_{2} & 1-c_{2} & -c_{1} \\
0 & 0 & 0 & 0 \\ \end{array}\right)\cr
M_{HIb}(b) =  \frac{1}{2}\left(\begin{array}{cccc}
0 &  c_{1} \left(1-\gamma\right) &  c_{1} \left(1-\gamma\right) & -c_{1}^2 \left(1-\gamma\right)^2 \\
0 & \left(1-\gamma\right) & \left(1-\gamma\right) & -c_{1} \left(1-\gamma\right)^2 \\
0 &  \left(1+\gamma\right) &\left(1+\gamma\right) & 4 c_{1} \kappa  \\
0 & 0 & 0 & 0 \\
\end{array}\right)\cr
M_{HIb}(c) = \frac{1}{2}\left(\begin{array}{cccc}
0 & 0 & 0 & 0 \\
0 & \left(1-\gamma\right) & (1-\gamma)^2(1+\gamma)^{-1} & 0 \\
0 & (1+\gamma)^2(1-\gamma)^{-1} & \left(\gamma+1\right) & 0 \\
0 & 0 & 0 & 0 \\
\end{array}\right) 
}    
\end{equation}
where $\gamma=\sqrt{4\kappa+1}$ .

\subsection{Extensions of known models}
\label{sec:knownmodels}
In this section, we look into models carrying asymmetric spin hoppings from our numerical results. We will group them for having similar nearest-neighbouring dynamics by decomposing each two-site $e_{i}$ in terms of $SU(2)$ spin matrices $S^{q}_{i}=\sigma^{q}_{i}/2$. We use the notations in \eref{eq:spinoprnotation} for brevity.
\begin{equation}
\label{eq:spinoprnotation}
\eqalign{
S_{i,xy}^{\pm} &= S_{i}^{x} \pm i S_{i}^{y} \cr
S_{i,xz}^{\pm} &= S_{i}^{x} \pm i S_{i}^{z} \cr
\mathcal{P}_{i,i+1} &= \frac{1}{2}\left(I + \sigma_{i}^{x}\sigma_{i+1}^{x} + \sigma_{i}^{y}\sigma_{i+1}^{y} + \sigma_{i}^{z}\sigma_{i+1}^{z}\right)
}
\end{equation}

We reiterate that our results arrive from the general treatment of solutions of the Yang Baxter equation. One may consider to map them into spinless fermions for interpreting them as Markov processes and is not solely limited by it. Hence they appear as non-hermitian spin chains for a broader consideration.

\subsubsection{Asymmetric hopping models}
$M_{BI}(a)$ from \eref{eq:BI} and $M_{HIa}(a)$ from \eref{eq:HIa} and $M_{HIb}(c)$ from \eref{eq:HIb} have similar dynamics with $M_{HIb}(a)$ from \eref{eq:HIb}, which we write in the following manner
\begin{equation}
\label{eq:ASEPext1}
\fl\eqalign{
\mathcal{M}_{i,i+1}=-\mathcal{K}_{i,i+1}(c_2) + \mathcal{Q}_{i,i+1} \cr \mathcal{K}_{i,i+1}(c_2) = c_{2}S_{i,xy}^{-}S_{i+1,xy}^{+}+(1-c_{2})S_{i,xy}^{+}S_{i+1,xy}^{-}+S_{i}^{z}S_{i+1}^{z}+\left(c_{2}-\frac{1}{2}\right)(S_{i+1}^{z}-S_{i}^{z})-\frac{1}{4} \cr \mathcal{Q}_{i,i+1}=S_{i}^{z}A_{i+1}^{-}-A_{i}^{-}S_{i+1}^{z} + \frac{1}{2}\left(A_{i}^{+}-A_{i+1}^{+}\right)\cr A_{i}^{\pm}=c_{1}S_{i,xy}^{+}\pm\left(\frac{\kappa+c_{2}(1-c_{2})}{c_{1}}\right)S_{i,xy}^{-}
}  
\end{equation}
where $\mathcal{K}_{i,i+1}$ resembles a spin-ASEP model. Additionally $\mathcal{Q}_{i,i+1}$ appears in the model and is written in various subparts. To understand the extra term, we will simplify the model by substituting $c_{1}=(c_{2}(c_{2}-1)-\kappa)^{1/2}$. Then 
\begin{equation}
\label{eq:ASEPext1Qmod}
\fl\mathcal{Q}_{i,i+1} = (\sqrt{c_{2}(c_{2}-1)-\kappa})\left[\frac{1}{2i}(S_{i,xz}^{+}S_{i+1,xz}^{-}-S_{i,xz}^{-}S_{i+1,xz}^{+})+i(S_{i}^{y}-S_{i+1}^{y}))\right] 
\end{equation}
which represents an (XZ-aligned) spin-chain with a hermiticity-breaking subterm. With $\kappa=c_2(c_2-1)$, the extra term vanishes and we recover the ASEP model. In this manner we identify \textit{extensions} on already studied models. 

Another such extension is given by models of $M_{BN}(a)$ from \eref{eq:BN}, $M_{HN}(b), M_{HN}(c)$ from \eref{eq:HN} and $M_{HIb}(b)$ from \eref{eq:HIb}. We write $M_{HIb}(b)$ as follows
\begin{equation}
\fl\eqalign{
&\mathcal{M}_{i,i+1}= \mathcal{K}_{i,i+1}\left(\frac{1+\gamma}{2}\right) + \mathcal{Q}_{i,i+1} \cr
&\mathcal{Q}_{i,i+1}=-c_{1}^{2}\frac{(1-\gamma)^{2}}{2}S_{xy,i}^{+}S_{xy,i+1}^{+}-\frac{1}{4} c_{1} \gamma(1-\gamma)\left(S_{xy,i+1}^{+}-S_{xy,i}^{+}\right) -2S_{i}^{z}S_{i+1}^{z}\cr &-\frac{1}{2} c_{1} \left(\gamma-1\right)\left(\gamma+2\right)S_{i}^{z}S_{xy,i+1}^{+} + \frac{1}{2} c_{1} \left(\gamma-1\right)\left(\gamma-2\right)S_{xy,i}^{+}S_{i+1}^{z} +\frac{1}{2}
}    
\end{equation}
The additional term possess the dynamics of spin-creation operations. Imposing $c_1=0$ recovers the model.

Finally we conclude that there are similar extensions to the TASEP (Totally ASEP) models, where the spin-hoppings are allowed in one direction in the periodic chain. They are models $M_{BI}(b), M_{BI}(d)$ from \eref{eq:BI} and $M_{HIa}(b), M_{HIa}(c)$ from \eref{eq:HIa}.

\subsubsection{Anti-Hermitian model} We find a model $M_{HN}(a)$ from \eref{eq:HN} which extends an anti-hermitian spin chain. $M_{HN}(a)$ is written as follows
\begin{equation}
\label{eq:antiASEPext1}
\fl\eqalign{
&\mathcal{M}_{i,i+1}=\left[c_{1}S_{i,xy}^{-}S_{i+1,xy}^{+}-c_{1}S_{i,xy}^{+}S_{i+1,xy}^{-}+c_{1}(S_{i+1}^{z}-S_{i}^{z})\right] + \mathcal{Q}_{i,i+1} \cr
&\mathcal{Q}_{i,i+1}=S_{i}^{z}A_{i+1}^{-}-A_{i}^{-}S_{i+1}^{z} + \frac{1}{2}\left(A_{i}^{+}-A_{i+1}^{+}\right)\cr
&A_{i}^{\pm}=\left(\left(\frac{\kappa-c_{1}^{2}}{c_{2}}\right)S_{i,xy}^{+}\pm c_{2}S_{i,xy}^{-}\right)
}    
\end{equation}

Upto relevant parameter substitutions, we can rewrite $\mathcal{Q}_{i,i+1}$ in \eref{eq:antiASEPext1} similar to that in \eref{eq:ASEPext1Qmod}. Both systems can be further combined together into the following rank-1 model
\begin{equation}
\eqalign{
\fl e_{i}=-\frac{p}{4}+\frac{2q-p}{2}\left(S_{i}^{z}-S_{i+1}^{z}\right)+\left(qS_{i,xy}^{+}S_{i+1,xy}^{-}+(p-q)S_{i,xy}^{-}S_{i+1,xy}^{+}\right)+p S_{i}^{z}S_{i+1}^{z}\cr 2s \left(S_{i}^{z}S_{i+1}^{x}-S_{i}^{x}S_{i+1}^{z}\right)+is\left(S_{i}^{y}-S_{i+1}^{y}\right)
}    
\end{equation}

satisfying the below TL algebra,
\begin{equation}
\fl\eqalign{
e_{i}e_{i+1}e_{i}=(s^2-q(q-p)) e_{i} \cr
e_{i+1}e_{i}e_{i+1}= (s^2-q(q-p))e_{i+1} \cr
e_{i}^{2}=-pe_{i}
}    
\end{equation}
which is also a solution of the YBE with the R-Matrix given from \eref{eq:Rmatr}.

\subsubsection{Spin-creation models} Finally we have $M_{BN}(b)$ from \eref{eq:BN} and $M_{BI}(c), M_{BI}(e)$ from \eref{eq:BI} which represents the model with sole spin-creation operations. For example, we write $M_{BI}(c)$ as follows 
\begin{equation}
\fl\eqalign{
\mathcal{M}_{i,i+1} &= c_{2}S_{i,xy}^{+}S_{i+1,xy}^{+}-S_{i}^{z}S_{i+1}^{z}+\frac{1}{2}(S_{i+1}^{z}-S_{i}^{z})+\frac{1}{4}+\mathcal{Q}_{i,i+1} \cr
\mathcal{Q}_{i,i+1} &= \frac{c_{2}}{c_{1}}S_{i,xy}^{+}S_{i+1}^{z}-c_{1}S_{i}^{z}S_{i+1,xy}^{+}+\frac{1}{2}\left(\frac{c_{2}}{c_{1}}S_{i,xy}^{+}+c_{1}S_{i+1,xy}^{+}\right)
}    
\end{equation}
This completes the summary of all rank-1 models that are obtained through the numerical analysis. We further comment that the free parameters in our models are not further reduced via similarity transformation, which may reveal already known integrable models. Hence we do not claim of finding new models. Nevertheless one is yet to investigate the problem of identifying higher rank models exhaustively.

\subsection{Higher rank models}
For completeness we also showcase some of the models $e_{i}$ constructed through the nilpotent and idempotent matrices of rank $r>1$ which we have managed to identify from our numerical analysis. For brevity,  $\lambda,\kappa$ and $\omega$ are written with respect to the model's parameters.

\subsubsection{Some rank 2 cases}
We have identified some Hecke-Idempotent models in \tref{table:rk2hi} where $e_{i} = -c_{\omega}(\lambda)\mathbf{B}_{2} + \nu_{+}\mathbf{I}$. They all satisfy the below subcase of \eref{eq:alg}
\begin{equation}
\eqalign{
    &e_{i}e_{i+1}e_{i}=e_{i+1}e_{i}e_{i+1} \neq 0, \cr
    &e_{i}^{2}=\lambda-\omega e_{i}.
}    
\end{equation}

\begin{table}[htpb]
\centering
\caption{\label{table:rk2hi}Some rank-2 Hecke-Idempotent models}
\begin{tabular}{@{}cccc@{}}
\br $\textrm{Model}$ & $e_{i}$ & $\omega$ & $\lambda$ \\ \mr
$B1$ &$\eqalign{-\frac{s}{2}+\left((p+s)S_{i}^{z}+pS_{i+1}^{z}\right) + q S_{i, xy}^{-}S_{i+1, xy}^{-} \cr + \left(p S_{i, xy}^{+}S_{i+1, xy}^{-} + (p+s) S_{i, xy}^{-}S_{i+1, xy}^{+}\right) }$ & $s$ & $p(p+s)$ \\ \mr
$B2$ &$\eqalign{-\left(S_{i+1}^{y}+S^{y}_{i}\right) + r S_{i, xz}^{-}S_{i+1, xz}^{-} +\left(S_{i, xz}^{-}S_{i+1, xz}^{+} + S_{i, xz}^{+}S_{i+1, xz}^{-}\right)}$  &      $0$    &      $1$     \\ \mr
$B3$ &$k\left(S_{i}^{z}+S_{i+1}^{z}\right) + \left(k^2 S_{i, xy}^{+}S_{i+1, xy}^{-} +  S_{i, xy}^{-}S_{i+1, xy}^{+}\right) $ & $0$ & $k^2$ \\ \br
\end{tabular}
\end{table}
It is interesting to note about model $B1$ that it becomes a Hecke-Nilpotent model of rank 2 when $s=-2p$.

\subsubsection{A rank 3 case} In the end, we finally provide one model of the form $e_{i} = -c_{\omega}(\lambda)\mathbf{B}_{3} + \nu_{+}\mathbf{I}$ which follows the below algebra
\begin{equation}
\eqalign{
    &e_{i}e_{i+1}e_{i}=t_{i,i+1}-\kappa e_{i}, \cr
    &e_{i+1}e_{i}e_{i+1}=t_{i,i+1}-\kappa e_{i+1}, \cr
    &e_{i}^{2}=\lambda;
}    
\end{equation}
and is given as below
\begin{equation}
\eqalign{
    &e_{i} = -r\mathcal{P}_{i,i+1} + 2\sqrt{p-r^2}\left(S_{i}^{z}S_{i+1}^{x}-S_{i}^{x}S_{i+1}^{z}\right) + i\sqrt{p-r^2}\left(S_{i+1}^{y}-S_{i}^{y}\right), \cr
    &t_{i,i+1} = (pr-2r^3) -r^2\left(e_{i}+e_{i+1}\right) -r \left\{e_{i},e_{i+1}\right\}, \cr
    &\kappa = r^2 - p, \cr
    &\lambda= r^2.
}   
\end{equation}
This model is also discussed within the setting of \cite{Gwa92_2} and presents into our classification as follows -- a rank-3 HI model when $p\neq 0$, a rank-3 BI model when $p=0$ and a rank-1 HN model when $r=0$. The case of $p=r^{2}$ represents the well known Heisenberg XXX model.

%% file: section-Xreps.tex
\section{Hubbard-type  representation of (degenerate) Hecke algebraic models}
In this section we are interested to bring note of a (degenerate) Hecke algebraic model within the representation of Hubbard X-operators which illuminates on the appearance of non-trivial spin-chain models from our numerical analysis. 

First we introduce the Hubbard operators as
\begin{equation}
\eqalign{X^{\alpha\beta}_{i}=(|\alpha\rangle\langle\beta|)_{i},\quad X^{\alpha\beta}_{i}X^{\lambda\gamma}_{i}=\delta^{\beta\lambda}X^{\alpha\gamma}_{i}\\\sum_{\alpha}X^{\alpha\alpha}_{i}=1,\quad [X^{\alpha\beta}_{i},X^{\delta\gamma}_{j}]_{\pm}=(\delta^{\beta\delta}X^{\alpha\gamma}_{i}\pm\delta^{\alpha\gamma}X_{i}^{\beta\gamma})\delta_{ij},}
\end{equation}

where $[A,B]_{\pm}=AB-(-1)^{p(A)p(B)}BA$ denotes a graded commutator with $p(A)$ being the fermionic parity of an operator $A$. We consider the particular choice of the basis for the operators
\begin{equation}
    |0\rangle=\left|\uparrow\downarrow\right\rangle, \quad |1\rangle=\left|\downarrow\right\rangle,\quad |2\rangle=\left|\uparrow\right\rangle, \quad |3\rangle=\left|\circ\right\rangle,
\end{equation}

where states $|0\rangle$ and $|3\rangle$ are bosonic whereas states $|1\rangle$ and $|2\rangle$ are fermionic. With the Greek indices running through integers 0 to 3, a particular representation of the $X$-operators is given by
\begin{equation}
\label{eq:Xrep}
[X^{\alpha\beta}_{i}] =  \left(\begin{array}{cccc}
n_{\downarrow}n_{\uparrow} & n_{\downarrow}c^{\dag}_{\uparrow} & -c_{\downarrow}^{\dag}n_{\uparrow} & c^{\dag}_{\uparrow}c^{\dag}_{\downarrow} \\
n_{\downarrow}c_{\uparrow} & n_{\downarrow}(1-n_{\uparrow}) & c_{\downarrow}^{\dag}c_{\uparrow} & c_{\downarrow}^{\dag}(1-n_{\uparrow}) \\
-c_{\downarrow}n_{\uparrow} & c_{\uparrow}^{\dag}c_{\downarrow} & (1-n_{\downarrow})n_{\uparrow} & c^{\dag}_{\uparrow}(1-n_{\downarrow} \\
c_{\downarrow}c_{\uparrow} & c_{\downarrow}(1-n_{\uparrow}) & c_{\uparrow}(1-n_{\downarrow}) & (1-n_{\uparrow})(1-n_{\downarrow})
\end{array}\right).
\end{equation}

Next by using the bond notation $O_{i,i+1}\equiv O_{i}$, we introduce the following set of operators
\begin{subequations}
\begin{equation}
\eqalign{a^{\dag}_{i}=X^{30}_{i}X^{30}_{i+1}-X^{10}_{i}X^{20}_{i+1}+X^{20}_{i}X^{10}_{i+1},\\    
a_{i}=X^{03}_{i}X^{03}_{i+1}+X^{01}_{i}X^{02}_{i+1}-X^{02}_{i}X^{01}_{i+1},}    
\end{equation}
\begin{equation}
b_{i}=\sum_{a}(-1)^{p(a)}X_{i}^{0a}X_{i+1}^{a0},\quad b^{\dag}_{i}=\sum_{a}X^{a0}_{i}X^{0a}_{i+1},    
\end{equation}
\begin{equation}
\eqalign{p_{i}^{0}=(1-X_{i}^{00})(1-X^{00}_{i+1}),\;\; p^{1}_{i}=X^{00}_{i}(1-X^{00}_{i+1}),\\ p_{i}^{2}=(1-X^{00}_{i})X^{00}_{i+1},\;\; p^{3}_{i}=X^{00}_{i}X^{00}_{i+1},}   
\end{equation}
\begin{equation}
B_{i}=\sum_{a,b}(-1)^{p(b)}X_{i}^{ab}X_{i+1}^{ba},\quad r_{i}=a^{\dag}_{i}a_{i},    
\end{equation}
\end{subequations}

where the Latin indices run from integers 1 to 3. These operators satisfy the following quasi-local algebra \cite{Gritsev_2003}
\begin{subequations}
\begin{equation}
r_{i}r_{i+1}r_{i}=r_{i}p_{i+1}^{0},\quad r_{i+1}r_{i}r_{i+1}=p_{i}^{0}r_{i+1},    
\end{equation}
\begin{equation}
B_{i}B_{i+1}B_{i}=B_{i+1}B_{i}B_{i+1},    
\end{equation}
\begin{equation}
b^{\dag}_{i}r_{i+1}b_{i}=b_{i+1}r_{i}b_{i+1}^{\dag},\quad b_{i}^{\dag}B_{i+1}b_{i}=b_{i+1}B_{i}b^{\dag}_{i+1},    
\end{equation}
\begin{equation}
b_{i}b_{i+1}a_{i}^{\dag}=a_{i+1}^{\dag}p_{i}^{3},\quad b_{i}a_{i+1}a_{i}^{\dag}=p_{i}^{1}b^{\dag}_{i+1},    
\end{equation}
\begin{equation}
b_{i}a_{i+1}r_{i}=b^{\dag}_{i+1}a_{i},\quad b_{i}a_{i+1}B_{i}=b^{\dag}_{i+1}a_{i}B_{i+1},    
\end{equation}
\begin{equation}
b_{i}b_{i+1}B_{i}=B_{i+1}b_{i}b_{i+1},    
\end{equation}
\begin{equation}
b^{\dag}_{i}a^{\dag}_{i+1}=b_{i+1}a_{i}^{\dag}    
\end{equation}
\end{subequations}

among the other relations. For a single bond, we have $\;b_{i}^{\dag}b_{i}=p_{i}^{2},\;b_{i}b_{i}^{\dag}=p_{i}^{1},\;B_{1}^{2}=p_{i}^{0}$ and for $|i-j|>1$ all the operators commute. 
One {\it particular} Baxterization of this algebra is provided by the operator 
\begin{equation}
e_{i}=-\left[b_{i}+b_{i}^{\dag}+\left(\alpha+\frac{1}{\alpha}\right)(p_{i}^{0}+p_{i}^{3})+\alpha p_{i}^{1}+\frac{1}{\alpha}p_{i}^{2}-\left(\alpha+\frac{1}{\alpha}\right)\right]
\end{equation}

which satisfies the following Hecke algebra relations \eref{eq:alg}
\begin{equation}
\eqalign{
e_{i}e_{i+1}e_{i}-e_{i}=e_{i+1}e_{i}e_{i+1}-e_{i+1},\\
e_{i}e_{j}=e_{j}e_{i},\qquad |i-j|>1,\\
e_{i}^{2}=\left(\alpha+\frac{1}{\alpha}\right)e_{i}.
}
\end{equation}

In the limit of $\alpha=-1/\alpha$ the relations reduces to the Temperley-Lieb algebra. For the limit of $\alpha=1$, one can write the generator $h_{i} = 1 - q_{i}$, where  $q_{i}=b_{i}+b_{i}^{\dag}+p_{i}^{0}+p_{i}^{3}$ further satisfy the braid equation \eref{eq:braidequiv}.

These algebraic structures correspond to a variation of an integrable $t-J$-type models
\begin{equation}
\eqalign{
H=-&\sum_{i,\sigma}(n_{i,\bar{\sigma}}c^{\dag}_{i,\sigma}c_{i+1,\sigma}n_{i,\bar{\sigma}}+n_{i+1,\bar{\sigma}}c^{\dag}_{i+1,\sigma}c_{i,\sigma}n_{i,\bar{\sigma}}\\ &+\eta^{+}_{i}\eta_{i+1}^{-}+\eta^{+}_{i+1}\eta_{i}^{-}+Vn_{i,\uparrow}n_{i,\downarrow}n_{i+1,\uparrow}n_{i+1,\downarrow}+Un_{i,\uparrow}n_{i,\downarrow})
}
\label{eq:susyHam}
\end{equation}

with arbitrary $V$ and $U$. Here $\eta^{+}_{i}=c^{\dag}_{i,\uparrow}c^{\dag}_{i,\downarrow}$, $\eta_{i}^{-}=c_{i,\downarrow}c_{i,\uparrow}$ are generators of the pairing $su(2)$ algebra
\begin{equation}
[\eta^{+}_{i},\eta^{-}_{i}]=2\eta^{z}_{i},\quad [\eta^{z}_{i},\eta^{\pm}_{i}]=2\eta^{z}_{i}, \quad 2\eta^{z}_{i}=n_{i,\uparrow}+n_{i,\downarrow}-1.
\end{equation}

We further note that the Hamiltonian \eref{eq:susyHam} commutes with the generators of the supersymmetric $su(2|1)$ algebra. Since in practice all our models in \sref{sec:knownmodels} may be written into fermionic operations through a pseudospin representation, we can tie our spin-chain models as a manifestation of integrable variations of the Hubbard model.

%% file: section-conclusion.tex
Motivated by the earlier papers on exactly solvable exclusion processes here we presented a number of (possibly) new solutions of the Yang-Baxter equation related to low-rank matrices and degenerate versions of Hecke-related algebraic structures. We also wrote spin-1/2 versions of the corresponding exclusion processes and showed its connection with integrable hubbard-type models. In the future we plan to examine their critical and dynamical properties. 

\ack{We would like to thank Professor Fabian Essler for useful comments and encouragements. Work of SB and VG is partially supported by the Delta Institute for Theoretical Physics (DITP). DITP consortium, a
program of the Netherlands Organization for Scientific
Research (NWO) is funded by the Dutch Ministry of Education, Culture and Science (OCW). VG is also partially supported by the Pauli Center for Theoretical Physics at the ETH Zurich. }

%% file: section-appendices.tex
\section{Calculation towards finding \texorpdfstring{$f(x,y)$}{f(x,y)}}
\label{appendix:A}
We first note the constraints on the function $f_{ij} \equiv f\left(u_i,u_j\right)$ 
\begin{equation}
\eqalign{
\lim_{y\rightarrow x}f_{x,y} = 0\cr
\frac{\left(f_{12}+f_{21}\right)}{f_{12}f_{21}} =\omega\cr
\frac{1}{f_{12}f_{13}f_{23}}\left(f_{12}+f_{23}-f_{13}-\omega f_{12}f_{23}\right) =\kappa 
}    
\end{equation}
where $f_{ij}$ is the following ansatz
\begin{equation}
\label{eq:ansatz2}
f(x,y)=\frac{x-y}{S(x,y)},\;\;S(x,y)=\sum_{i,j=0}^{N}d_{ij}x^{i}y^{j}.    
\end{equation}
It immediately satisfies $\lim_{y\rightarrow x}f(x,y)=0$. Substituting the ansatz to the second constraint reveals 
\begin{equation}
\sum_{i,j=0,i\neq j}^{N}(d_{ij}-d_{ji})x^{i}y^{j}=\omega(x-y)
\end{equation}
which resolves by identifying
\begin{equation}
\eqalign{
d_{10}=d_{01}+\omega\cr
d_{ij}= d_{ji},\;\;i \neq j,\;\; (i,j)\not\in\{(1,0),(0,1)\}
}    
\end{equation}
Expanding the third constraint gets tedious if all variables $(u_1,u_2,u_3)$ are considered. Hence we consider $u_2=0$ and expand as follows
\begin{equation}
\eqalign{
f(u_1,0)+f(0,u_3)-f(u_1,u_3)-\omega f(u_1,0)f(0,u_3)\cr -\kappa f(u_1,0)f(u_1,u_3)f(0,u_3)=0} 
\end{equation}

which after substituting the form of $f(x,y)$, becomes
\begin{equation}
\label{eq:cond2f}
\eqalign{
S(u_1,u_3)(u_1 S(0,u_3)-u_3 S(u_1,0)+\omega u_1 u_3)\cr -(u_1-u_3)(S(u_1,0)S(0,u_3)+\kappa u_1 u_3)=0}    
\end{equation}

\subsubsection*{Expansions}
The numerator terms of \eref{eq:cond2f} are expanded term-wise as follows
\begin{equation}
\eqalign{
u_1 S(0,u_3) S(u_1,u_3) &= u_1\left(\sum_{i=0}^{N}d_{0,i}u_{3}^{i}\right)\left(\sum_{j,k=0}^{N}d_{j,k}u_{1}^{j}u_{3}^{k}\right) \cr
&= \sum_{i,j,k=0}^{N}d_{0,i}d_{j,k}u_{1}^{j+1}u_{3}^{i+k} \cr
&= \sum_{j=0}^{N}\sum_{n=0}^{2N}\left(\sum_{n=i+k}^{0\leq i,k \leq N}d_{0,i}d_{j,k}u_{3}^{n}\right)u_{1}^{j+1} \cr
&= \sum_{i=0}^{N}\sum_{j=0}^{2N}\left(\sum_{j=\alpha+\beta}^{0\leq \alpha,\beta \leq N}d_{0,\alpha}d_{i,\beta}\right)u_{3}^{j}u_{1}^{i+1} \cr
}    
\end{equation}
\begin{equation}
\eqalign{
u_3 S(u_1,0) S(u_1,u_3) &= u_3\left(\sum_{i=0}^{N}d_{i,0}u_{1}^{i}\right)\left(\sum_{j,k=0}^{N}d_{j,k}u_{1}^{j}u_{3}^{k}\right) \cr
&= \sum_{i,j,k=0}^{N}d_{i,0}d_{j,k}u_{1}^{i+j}u_{3}^{k+1} \cr
&= \sum_{k=0}^{N}\sum_{n=0}^{2N}\left(\sum_{n=i+j}^{0\leq i,j \leq N}d_{i,0}d_{j,k}u_{1}^{n}\right)u_{3}^{k+1} \cr
&= \sum_{j=0}^{N}\sum_{i=0}^{2N}\left(\sum_{i=\alpha+\beta}^{0\leq \alpha,\beta \leq N}d_{\alpha,0}d_{\beta,j}\right)u_{1}^{i}u_{3}^{j+1} \cr
}    
\end{equation}
\begin{equation}
\omega u_1 u_3 S(u_1, u_3) = \omega \sum_{i,j=0}^{N} d_{i,j}u_{1}^{i+1}u_{3}^{j+1}
\end{equation}
\begin{equation}
\eqalign{
(u_1-u_3)S(u_1,0)S(0,u_3) &= (u_1-u_3)\sum_{i,j=0}^{N}(d_{i,0}d_{0,j}u_{1}^{i}u_{3}^{j}) \cr
&= \sum_{i,j=0}^{N}d_{i,0}d_{0,j}(u_{1}^{i+1}u_{3}^{j}-u_{1}^{i}u_{3}^{j+1})
}    
\end{equation}
\begin{equation}
\kappa u_{1}u_{3}(u_{1}-u_{3})=\kappa(u_{1}^{2}u_{3}-u_{1}u_{3}^{2})
\end{equation}

\subsubsection*{Cases}

We now consider looking into the pre-factors of each monomial terms of the numerator of \eref{eq:cond2f} and equate it to zero. 
\begin{enumerate}
\item $u_{1}^{i}u_{3}^{j+1},\;i\geq N+2,\;\;0\leq j\leq N$ : 
\begin{equation}
\fl\sum_{i=\alpha+\beta}^{0\leq \alpha,\beta \leq N}(-d_{\alpha,0}d_{\beta,j})=0
\end{equation}
We start by taking $i=2N$, the maximal power possible in this case and find $d_{N,0}d_{N,j}=0$. By putting $d_{N,0}=0$ for all values of $j$, we can proceed with $i=2N-1$ and impose $d_{N-1,0}=0$ similarly. In this way, we impose $d_{i,0}=0,\;2\leq i \leq N$.

\item $u_{1}^{i+1}u_{3}^{j},\;j\geq N+2,\;\;0\leq i\leq N$ : 
\begin{equation}
\fl\sum_{j=\alpha+\beta}^{0\leq \alpha,\beta \leq N}d_{0,\alpha}d_{i,\beta}=0.    
\end{equation}
We impose $d_{i,j}=d_{j,i}$ for all $i\neq j$ except for $(i,j)\in\{(1,0),(0,1)\}$ to use the previous remark in keeping it zero.

\item $u_{1}^{N+1}u_{3}^{N+1}$ : 
\begin{equation}
\fl(d_{0,1}d_{N,N}-d_{1,0}d_{N,N}+\omega d_{N,N})=0
\end{equation}
which is satisfied by the earlier result that $d_{1,0}=d_{0,1}+\omega$.

\item $u_{1}^{N+1}u_{3}^{j},\;1\leq j\leq N$ : 
\begin{equation}
\fl\eqalign{
&\sum_{j=\alpha+\beta}^{0\leq \alpha,\beta \leq N}(d_{0,\alpha}d_{N,\beta})-\sum_{N+1=\alpha+\beta}^{0\leq \alpha,\beta \leq N}(d_{\alpha,0}d_{\beta,j-1})-d_{N,0}d_{0,j}+\omega d_{N,j-1} \\
&=d_{0,0}d_{N,j}+d_{0,1}d_{N,j-1}-d_{1,0}d_{N,j-1}+\omega d_{N,j-1} \\
&=d_{0,0}d_{N,j} = 0
}
\end{equation}
If $d_{0,0}=0$, then $f(u,u)$ will be indeterminate. It can be null for $d_{N,j}=0$ for all $j$. 

For the  $u_{1}^{N+1}$ term, the prefactor is $d_{0,0}d_{N,0}-d_{N,0}d_{0,0}=0$.

For $u_{1}^{i}u_{3}^{N+1},\;1\leq i\leq N$,
\begin{equation}
\fl\eqalign{
&\sum_{N+1=\alpha+\beta}^{0\leq \alpha,\beta \leq N}(d_{0,\alpha}d_{i-1,\beta})-\sum_{i=\alpha+\beta}^{0\leq \alpha,\beta \leq N}(d_{\alpha,0}d_{\beta,N})+d_{i,0}d_{0,N}+\omega d_{N,j-1} \\
&=d_{0,1}d_{i-1,N}-d_{0,0}d_{i,N}-d_{1,0}d_{i-1,N}+\omega d_{N,j-1} \\
&=-d_{0,0}d_{i,N} = 0
}
\end{equation}
For $u_3^{N+1}$, the prefactor is $-d_{0,0}d_{0,N}+d_{0,0}d_{0,N}=0$.

\item $u_{1}^{p}u_{3}^{q},\;2\leq p,q \leq N$ : 
\begin{equation}
\fl\eqalign{
&\sum_{p=\alpha+\beta}^{0\leq \alpha,\beta \leq N}(d_{0,\alpha}d_{p-1,\beta}) - \sum_{q=\alpha+\beta}^{0\leq \alpha,\beta \leq N}(d_{\alpha,0}d_{\beta,q-1}) - (d_{p-1,0}d_{0,q}-d_{p,0}d_{0,q-1})+\omega d_{p-1,q-1} \\
&=d_{0,0}(d_{p-1,q}-d_{p,q-1})+d_{p-1,q-1}(d_{0,1}+\omega-d_{1,0}) \\
&=d_{0,0}(d_{p-1,q}-d_{p,q-1})=0
}    
\end{equation}
which is satisfied by putting $d_{p-1,q}=d_{p,q-1}$.

\item $u_{1}u_{3}^{q},\;3\leq q \leq N$ :
\begin{equation}
\fl\eqalign{
&\sum_{q=\alpha+\beta}^{0\leq \alpha,\beta \leq N}(d_{0,\alpha}d_{0,\beta}) - \sum_{1=\alpha+\beta}^{0\leq \alpha,\beta \leq N}(d_{\alpha,0}d_{\beta,q-1}) - (d_{0,0}d_{0,q}-d_{1,0}d_{0,q-1})+\omega d_{0,q-1} \\
&=d_{0,0}(d_{0,q}-d_{1,q-1})+d_{0,q-1}(d_{0,1}+\omega-d_{1,0}) \\
&=-d_{0,0}(d_{1,q-1})=0
}    
\end{equation}

which is satisfied by putting $d_{1,q-1}=0$ for all $q=3,\dots,N$. 

Similarly for $u_{1}^{p}u_{3},\;3\leq p \leq N$, the prefactor is
\begin{equation}
\fl\eqalign{
&\sum_{1=\alpha+\beta}^{0\leq \alpha,\beta \leq N}(d_{0,\alpha}d_{p-1,\beta}) -\sum_{p=\alpha+\beta}^{0\leq \alpha,\beta \leq N}(d_{\alpha,0}d_{\beta,0}) - (d_{p-1,0}d_{0,1}-d_{p,0}d_{0,0})+\omega d_{p-1,0} \\
&=d_{0,0}(d_{p-1,1}-d_{p,0})+d_{p-1,0}(d_{0,1}+\omega-d_{1,0}) \\
&=d_{0,0}d_{p-1,1} = 0
}    
\end{equation}

\item $u_{1}^{2}u_{3}$ : 
\begin{equation}
\fl\eqalign{
&\sum_{1=\alpha+\beta}^{0\leq \alpha,\beta \leq N}(d_{0,\alpha}d_{1,\beta}) -\sum_{2=\alpha+\beta}^{0\leq \alpha,\beta \leq N}(d_{\alpha,0}d_{\beta,0}) - (d_{1,0}d_{0,1}-d_{2,0}d_{0,0})+\omega d_{1,0}+\kappa \\
&=d_{0,0}(d_{1,1}-d_{2,0})+d_{1,0}(d_{0,1}+\omega-d_{1,0})-d_{1,0}d_{0,1}+\kappa \\
&=d_{0,0}d_{1,1}-d_{1,0}d_{0,1}+\kappa=0
}    
\end{equation}

Similarly for $u_{1}u_{3}^{2}$, the prefactor is
\begin{equation}
\fl\eqalign{
&\sum_{2=\alpha+\beta}^{0\leq \alpha,\beta \leq N}d_{0,\alpha}d_{0,\beta} - \sum_{1=\alpha+\beta}^{0\leq \alpha,\beta \leq N}d_{\alpha,0}d_{\beta,1} - (d_{0,0}d_{0,2}-d_{1,0}d_{0,1})+\omega d_{0,1} -\kappa\\
&=d_{0,0}(d_{0,2}-d_{1,1})+d_{0,1}(d_{0,1}+\omega-d_{1,0})+d_{1,0}d_{0,1}-\kappa \\
&=-(d_{0,0}d_{1,1}-d_{1,0}d_{0,1}+\kappa)=0
}    
\end{equation}

For both cases, it requires
\begin{equation}
\fl d_{1,1}=\frac{d_{1,0}d_{0,1}-\kappa}{d_{0,0}}
\end{equation}
\item $u_{1}u_{3}$ :
\begin{equation}
\fl\eqalign{
&\sum_{1=\alpha+\beta}^{0\leq \alpha,\beta \leq N}(d_{0,\alpha}d_{0,\beta}) - \sum_{1=\alpha+\beta}^{0\leq \alpha,\beta \leq N}(d_{\alpha,0}d_{\beta,0}) - (d_{0,0}d_{0,1}-d_{1,0}d_{0,0})+\omega d_{0,0} \\
&=d_{0,0}(d_{0,1}-d_{1,0})+d_{0,0}(d_{0,1}+\omega-d_{1,0})-d_{0,0}(d_{0,1}-d_{1,0}) \\
&=0
}    
\end{equation}

For $u_{1}^{p},\;1\leq p \leq N$, the prefactor is $d_{0,0}d_{p-1,0}-d_{p-1,0}d_{0,0}=0$. 

For $u_{3}^{q},\;1\leq q \leq N$, the prefactor is $-d_{0,0}d_{0,q-1}+d_{0,0}d_{0,q-1}=0$.

\end{enumerate}
Consolidating the conditions obtained for nullifying each sub-terms, we have
\begin{itemize}
\item[1.] $d_{1,0}=d_{0,1}+\omega$
\item[2.] $d_{1,1}=(d_{1,0}d_{0,1}-\kappa)/d_{0,0}$
\item[3.] $d_{i,0}=d_{0,i}=0,\;2\leq i \leq N$
\item[4.] $d_{j,N}=d_{N,j}=0,\;1\leq j \leq N$
\item[5.] $d_{1,p}=d_{p,1}=0,\;2\leq p \leq N-1$
\item[6.] $d_{p-1,q}=d_{p,q-1},\;2\leq p,q\leq N$
\end{itemize}
We use the condition 4, 5 and 6 in the list to show that except for $d_{0,0},d_{1,1},d_{0,1}$ and $d_{1,0}$, all values of $d_{i,j}$ are 0. We demonstrate this by using $d_{p-1,q}=d_{p,q-1}$ as follows -
\begin{equation}
\label{eq:equalchain}
\fl\eqalign{
0&=d_{j,N}=d_{j+1,N-1}=d_{j+2,N-2}=\dots=d_{N,j},\;\;1\leq j\leq N \cr
0&=d_{1,j}=d_{2,j-1}=d_{3,j-2}=\dots=d_{j,1},\;\;2\leq j\leq N-1
}    
\end{equation}

We represent \eref{eq:equalchain} in a diagrammatic way by representing $d_{i,j}$ as $(i,j)$, and indicate the nullification as follows (for $N=4$ as an example)

\begin{center}
\includegraphics[scale=1.4]{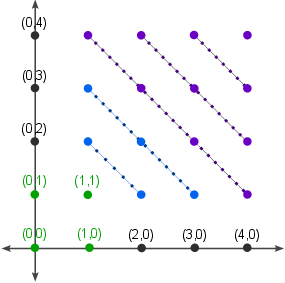}
\end{center}
The green points represents the coefficients which are non-zero. 

The ansatz \eref{eq:ansatz2} then becomes
\begin{equation}
\fl f(x,y)=\frac{x-y}{d_{0,0}+d_{0,1}(x+y)+\omega x+d_{1,1}xy},\;\;d_{1,1}=\frac{(d_{0,1}+\omega)d_{0,1}-\kappa}{d_{0,0}}
\end{equation}
and also satisfies
\begin{equation}
\eqalign{f(u_1,u_2)+f(u_2,u_3)-f(u_1,u_3)-\omega f(u_1,u_2)f(u_2,u_3)\cr-\kappa f(u_1,u_2)f(u_1,u_3)f(u_2,u_3)=0.}
\end{equation}

With the given form of $f(x,y)$, we substitute $d_{0,0}=c_{0}^{2}$ and $d_{0,1}=c_{0}c_{1}$ to rewrite it as
\begin{equation}
\fl f(x,y)=\frac{x-y}{c_0^{2}+c_{0}c_{1}(x+y)+c_{1}^{2}xy+\omega x+\left(\frac{c_1}{c_0}\omega-\frac{\kappa}{c_{0}^{2}}\right)xy}
\end{equation}

\section{Idempotent and degree-2 Nilpotent matrices of rank \texorpdfstring{$r$}{r}}
\label{appendix:B}
A rank $r$ square matrix of dimension $D$ is
\begin{equation}
\label{eq:BNform}
    \mathbf{A}_{r} = \sum_{i=1}^{r}C_{i}X_{i}^{T}
\end{equation}
where $\{C_{i}|\;1\leq i\leq r\}$ and $\{X_{i}|\;1\leq i \leq r\}$ are sets of $D$-dimensional linearly independent column vectors. 

To construct an idempotent matrix $\mathbf{B}_{r}$ of rank $r$, we require its Jordan canonical form to be a diagonal matrix with $r$ entries of $1$ and $0$ for the rest (in any order). Hence it can be written, upto similarity, in the following form
\begin{equation}
    \mathbf{B}_{r} = \mathbf{Q}\;\textrm{diag}[\underbrace{1,\dots,1}_{r},0,\dots,0]\;\mathbf{Q}^{-1}
\end{equation}
where $Q$ is any general invertible $D\times D$ matrix. By using 
\begin{equation}
    E_{i} = (0,\dots,\underbrace{1}_{i},0,\dots,0)^{T}
\end{equation}
where $1$ is in the $i$-th position of the column vector $E_{i}$, we can write $\mathbf{B}_{r}$ as
\begin{equation}
    \mathbf{B}_{r} = \sum_{i=1}^{r} \mathbf{Q}E_{i}E_{i}^{T}\mathbf{Q}^{-1} = \sum_{i=1}^{r}C_{i}X_{i}^{T}
\end{equation}
where $C_{i}=\mathbf{Q}E_{i}$ and $X_{i}^{T}=E_{i}^{T}\mathbf{Q}^{-1}$. Then we have
\begin{equation}
\label{eq:idcond}
   X_{i}^{T}C_{j} = \delta_{ij} 
\end{equation}
for any $\mathbf{A}_{r}$ to be idempotent. 

Similarly, to construct a Nilpotent matrix $\mathbf{N}_{r}$ of rank $r$ of degree $2$, i.e. $\mathbf{N}_{r}^{2}=\mathbf{0}$, we identify the following Jordan normal form (upto similarity)
\begin{equation}
\mathbf{Q}^{-1}\mathbf{N}_{r}\mathbf{Q}=\left[\begin{array}{cccccc}
S_1 & 0 & \ldots & 0 & \ldots & 0\\
0 & S_2 & \ldots & 0 & \ldots & 0\\
\vdots & \vdots & \ddots & \vdots & \ldots & 0 \\
0 & 0 & \ldots & S_r &\ldots & 0 \\
0 & 0 & \ldots & 0 &\ddots & 0 \\
0 & 0 & \ldots & 0 &\ldots & 0
\end{array}\right]
\end{equation}
where 
\begin{equation}
    S_{i} = \left[\begin{array}{cc}
        0 & 1 \\
        0 & 0
    \end{array}\right]
\end{equation}
and rest of the diagonal blocks are null. In general, $S_{i}$ along the diagonal blocks can be in any order. Notice that the rank of the nilpotent matrix satisfies $2r\leq D$. Writing the Jordan matrix in terms of $E_{i}$, we have
\begin{equation}
    \mathbf{N}_{r} = \sum_{i=1}^{r}\mathbf{Q}E_{2i-1}E_{2i}^{T}\mathbf{Q}^{-1} = \sum_{i=1}^{r} C_{i}X_{i}^{T}
\end{equation}
where $C_{i} = \mathbf{Q}E_{2i-1}$ and $X_{i}^{T} = E_{2i}^{T}\mathbf{Q}^{-1}$. Then we have
\begin{equation}
\label{eq:nlcond}
    X_{i}^{T}C_{j} = 0,\;\forall i,j
\end{equation}
for any $\mathbf{A}_{r}$ to be a nilpotent matrix of degree 2. 

\section{Symmetries of the R-Matrix}
\label{appendix:C}
The YBE is also an over-determined system of atmost cubic polynomials for solving the matrix elements of $\mathcal{R}(f(u,v))$, which resides in $\mathcal{A}\otimes\mathcal{A}$. We define a algebra homomorphism $\phi_{ij} : \mathcal{A}\otimes\mathcal{A}\rightarrow\mathcal{A}\otimes\mathcal{A}\otimes\mathcal{A}$ where
\begin{equation}
\label{eq:alghomor}
\eqalign{
\phi_{12}(x\otimes y) &= a\otimes b\otimes 1 \cr
\phi_{23}(x\otimes y) &= 1\otimes a\otimes b \cr
\phi_{13}(x\otimes y) &= a\otimes 1\otimes b \cr
}
\end{equation}
such that $\mathcal{R}_{ij} = \phi_{ij}(\mathcal{R})$. In this manner \eref{eq:RRR} is constructed.

By considering $\mathcal{A}\equiv\mathbf{C}^{2}$, the R-matrix becomes a $4\times4$ matrix. Then using \eref{eq:RRR}, we construct a maximal set of 64 equations with a total unknown of 16 variables. Using the notation followed in \cite{hietarinta} for the YBE equations of a $N^{2}\times N^{2}$ R-matrix, where $N=\dim(\mathcal{A})$ : 
\begin{equation}
\label{eq:HietarintaNotation}
\sum R_{ij}^{kl}E_{jl}\otimes E_{ik},\;\;E_{ij} = [(\delta_{ai}\delta_{bj})],\;{a,b\in\;\{1,2,\ldots,N^{2}\}},
\end{equation} 
which we will call it the Hietarinta notation, the YBE is written in the index notation
\begin{equation}
\label{eq:RRRHietarinta}
\fl \mathcal{R}_{j_{1} j_{2}}^{k_{1} k_{2}}(u_{1},u_{2}) \mathcal{R}_{k_{1} j_{3}}^{l_{1} k_{3}}(u_{1},u_{3}) \mathcal{R}_{k_{2} k_{3}}^{l_{2} l_{3}}(u_{2},u_{3})=\mathcal{R}_{j_{2} j_{3}}^{k_{2} k_{3}}(u_{2},u_{3}) \mathcal{R}_{j_{1} k_{3}}^{k_{1} l_{3}}(u_{1},u_{3}) \mathcal{R}_{k_{1} k_{2}}^{l_{1} l_{2}}(u_{1},u_{2})
\end{equation}
where repeated indices implies summation. The following form of the equation reveals the essential symmetries on the R-matrix, which are

\begin{enumerate}
\label{eq:Rsymms}
\item $\mathcal{R}_{ij}^{kl} \rightarrow \mathcal{R}_{kl}^{ij}$ \hspace{4.15cm}[Transposition]
\item $\mathcal{R}_{ij}^{kl} \rightarrow \mathcal{R}_{(i+n)\textrm{\scriptsize\,mod\,}N,\;(j+n)\textrm{\scriptsize\,mod\,}N}^{(k+n)\textrm{\scriptsize\,mod\,}N,\;(l+n)\textrm{\scriptsize\,mod\,}N}$ \hspace{0.7cm} [Index incremention]
\item $\mathcal{R}_{ij}^{kl} \rightarrow \mathcal{R}_{ji}^{lk}$\hspace{4.3cm}[Inversions]
\end{enumerate}
along with the local basis transformation and multiplicity freedom of the R-matrix
\begin{equation}
\mathcal{R} \rightarrow g(K\otimes K)\mathcal{R}(K \otimes K)^{-1}.
\end{equation}
for some non-singular $K\in\mathcal{A}$ and complex function $g$. These invariances allow in identifying redundant solutions.

\newpage
\section{Pseudocodes towards removing repeated CYBE solutions}
\label{appendix:D}
\subsection{Algorithm workflow}

To remove repeating R-Matrix solutions, we utilise their symmetries. We refer to Transposition, Inversions and Index incrementions as \textit{point transformations} and employ similarity transformations separately. The following workflow demonstrates the algorithm
\begin{center}
\includegraphics[scale=1]{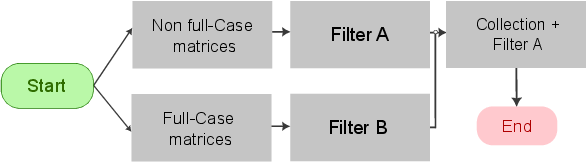}
\end{center}

\subsubsection*{Filter A workflow}
We consider all the matrix results $Rl$ which have some zero elements and generate equivalence classes \verb|caseunion[i]|$\equiv [i]$ based on the point transformations. Correspondingly, we construct a subcase graph \verb|subgraphs[i]|$\equiv g[i]$ for every $[i]$. 

The subcase graph is defined as $g[i]=\{a\rightarrow b \textrm{ if } a \textrm{ is transformable from }b\;\; \forall\;a,b\;\in[i]\}$. For checking if $b$ be transformed to $a$, first we transform both matrices closer to a triangular matrix via point transformations by using algorithm \ref{algr:Triformat}. Then we solve for the re-substitution of parameters in $b$ towards $a$ through algorithm \ref{algr:IsTransformable}. 

The pseudocode (algorithm \ref{algr:nfmodelclassifer}) is the main routine for generating the classifier objects. Algorithm \ref{algr:RMatrixInvariances} produce all possible permutations of matrices invariant under Transposition, Inversions and Index incrementions.

\begin{figure}[htbp]
    \centering
    \begin{minipage}{0.3\textwidth}
        \centering
        \includegraphics[width=1\textwidth]{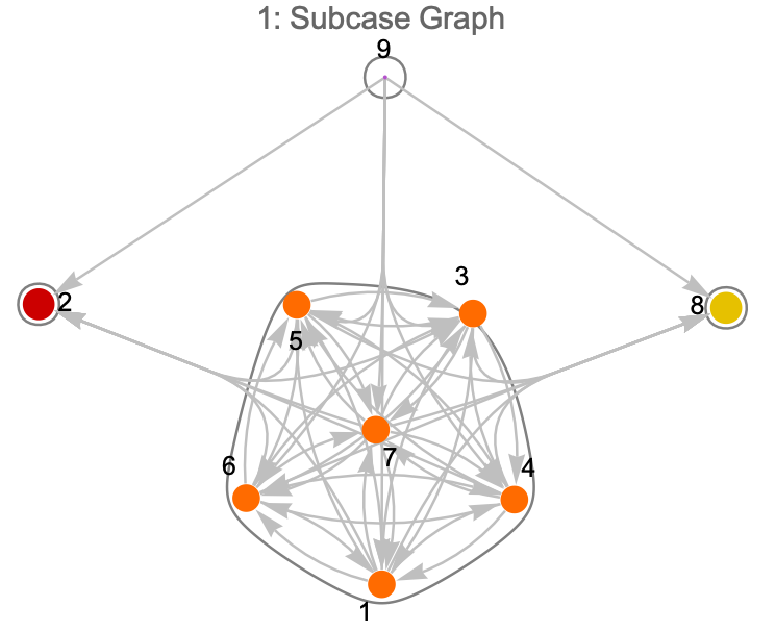} 
    \end{minipage}
    \begin{minipage}{0.3\textwidth}
        \centering
        \includegraphics[width=0.7\textwidth]{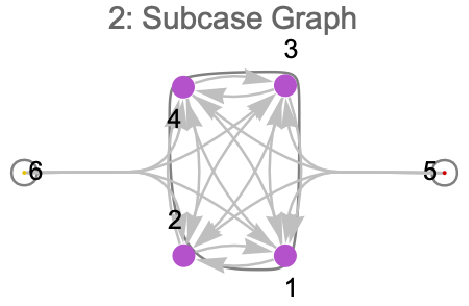} 
    \end{minipage}
    \caption{\label{fig:subgraphs} Examples of subcase graphs found in our numerical analysis. \\ The solution index $2,8$ and any of $1,2,3,4$ are considered from the $[i]$s \\ represented from left figure respectively.}
\end{figure}

For every corresponding $[i]$ we use the generated $g[i]$ to manually choose the results which are not subcases to other solutions. We show some of the generated $g[i]$s in \fref{fig:subgraphs}. In general, we circumvent highly parameterised results from our computation to consider simpler results for similarity transformations.

The union $U$ of all the selected results from every $[i]$ are then finally used to generate the following subcase graph $gs=\{b\rightarrow a \textrm{ if } f(a,b,t)\,\forall\,a,b\in U\}$, where $f(a,b,t)\equiv\,$\verb|IsSimilar[a,b,t]| from algorithm \ref{algr:IsSimilar}. $t$ is the parameter used to limit the computation time (in seconds). We manually remove the subcases and finalise the solution list.

\subsubsection*{Filter B workflow}
All the full-case matrices $Rf$ have the structure matrix which have no zero elements. They correspond to heavily coupled models which are untenable for our study. Hence we decided to break them into many non-full matrices to identify if there are any new results for further simplification.

First we generate the subcase graph $g[i]=\{a\rightarrow b \textrm{ if } a \textrm{ is transformable from }b\;\; \forall\;a,b\;\in Rf\}$. The results which are not the subcases of other solutions are taken. Then we use algorithm \ref{algr:FullCaseSimplify} to break them into valid matrix solutions having zero elements and finalise the list.

\subsubsection*{Collection + Filter A workflow}
From Filter A and Filter B workflow, all the results are collated and then run through the Filter A process again. It finally provide the final set of the unique solutions. In the context we are identifying $\mathcal{M}$ from \eref{eq:Rmatr}.

\subsection{Pseudocodes for the algorithm/routines used}

\begin{algorithm}[H]
\small
\caption{Non-full model classifier \label{algr:nfmodelclassifer}}
\KwDesc{Classifies all the non-full matrix results}
\KwInput{List of matrix results $Rl$}
\KwOutput{List \Ids{caseunion}, List \Ids{subgraphs}}
\KwRequire{Elements $N$ of $Rl$ are square matrices and $S(N)$ has some zero elements}
\BlankLine
\procedure(\procedurename{FullCaseSimplify}{$Rl$}){
\State{Local $\Ids{struct}\gets \{S(x)\;\forall\;x\in Rl\}$ (after removing duplicates)}
\State{Local $\Ids{matrsolgraph[i]} \gets \{x \in Rl \textrm{ s.t } S(x)=S(y)\}$ where $y$ is $i$-th element of \Ids{struct}} 
\State{Local $\Ids{grpcases}\gets \{1,2,\dots,\Call{Length}{struct}\}/\sim$ where $i\sim j$ if $S(i)\in\Call{RmatrixInvariances}{S(j),20}$}
\State{Local $\Ids{caseunion}[i]\gets\bigcup_{j\in [i]}\Ids{matrsolgraph[j]}\;\forall\;[i]\in \Ids{grpcases}$}
\BlankLine
\ForAll{$\Ids{caseunion[i]}\equiv x$}{
\State{$x \gets \{\Call{TriFormat}{i,120},\;i\in x\}$}
\State{Local $\Ids{subgraphs[i]} \gets \{\textrm{If }\Call{IsTransformable}{x_{n}, x_{m}, 120}\textrm{ then }m\rightarrow n,\forall \,1\leq n,m\leq \#x\}$}
\State{$\Ids{subgraphs[i]} \gets \Ids{subgraphs[i]}\textrm{ with all cliques identified }$}
}
\BlankLine
\State{$\Ids{caseunion}\gets\{\Ids{caseunion[i]},\;1\leq i \leq \Call{Length}{\Ids{grpcases}}\}$}
\State{$\Ids{subgraphs}\gets\{\Ids{subgraphs[i]},\;1\leq i \leq \Call{Length}{\Ids{grpcases}}\}$}
\State{Return \Ids{caseunion}, \Ids{subgraphs}}
}    
\end{algorithm}

\begin{algorithm}[H]
\small
\caption{Full model simplification routine \label{algr:FullCaseSimplify}}
\KwDesc{Breaks list of full-matrix results into non-full matrix results}
\KwInput{List of matrices $M$}
\KwOutput{List of non-full matrices \Ids{Ml}}
\KwRequire{Each matrix from $M$ are square matrices}
\BlankLine
\procedure(\procedurename{FullCaseSimplify}{$Ml$}){
\State{Local $\Ids{Ml}\gets \{\}$}
\ForEach{$m \in Ml$}{
\State{Local $\Ids{vars}\gets \textrm{ all variables of }m$}
\State{Local $\Ids{varsubs}\gets \textrm{ Variable replacement of }\Ids{vars}\textrm{ to unique }c_{i}$}
\State{$m \gets m \textrm{ after applying }\Ids{varsubs}$}
\State{Local $\Ids{temp}\gets \{m \textrm{ with } c_{i}=0 \forall c_{i}\}$}
\State{$\Ids{temp} \gets \Ids{temp} \textrm{ after removing Indeterminate cases}$}
\State{$\Ids{Ml} \gets \Ids{Ml}\cup \Ids{temp}$}
}
\State{$\textrm{Return }\Ids{Ml}$}
}
\end{algorithm}

\begin{algorithm}[H]
\small
\caption{Routine to make a table of R-Matrices invariant under its symmetries\label{algr:RMatrixInvariances}}
\KwDesc{Produce a set of matrices invariant of R-Matrix symmetries (except similarity transformation}
\KwInput{Matrix $R$, Integer $N$}
\KwOutput{List \Ids{RMatrices}}
\KwRequire{$N>0$, $R$ is a square matrix}
\BlankLine
\procedure(\procedurename{RMatrixInvariances}{$R,\;N$}){
\State{Local $\Ids{RMatrices} \gets \{R\}$}
\State{Local $\Ids{NewCases} \gets \{\}$}
\For{$i=1,\;i<N+1,\;\mathrel{i++}$}{
\State{$\Ids{NewCases} \gets \{\}$}
\State{$\Ids{NewCases} \gets \Ids{NewCases}\cup (\Ids{RMatrices} \textrm{\,after transposition})$} 
\State{$\Ids{NewCases} \gets \Ids{NewCases}\cup (\Ids{RMatrices} \textrm{\,after index incremention})$} 
\State{$\Ids{NewCases} \gets \Ids{NewCases}\cup (\Ids{RMatrices} \textrm{\,after inversion})$} 
\State{$\Ids{NewCases} \gets \Call{DeleteDuplicates}{\Ids{NewCases}}$} 
\ForEach{$m \in \Ids{NewCases}$}{
\If{$m \notin \Ids{RMatrices}$ }{
   \State $\Ids{RMatrices} \gets \Ids{RMatrices}\cup\{m\}$
}}}
\State{$\Ids{RMatrices} \gets \Call{DeleteDuplicates}{\Ids{RMatrices}}$}
\State{$\textrm{Return }\Ids{RMatrices}$}
}
\end{algorithm}

\begin{algorithm}[H]
\small
\caption{Routine to check if one R-Matrix be transformed to another one \label{algr:IsTransformable}}
\KwDesc{Checks if constant matrix $M_1$ be transformed to $M_2$ by variable substitution}
\KwInput{Matrix $M_1, M_2$, Integer $t$ (time, seconds)}
\KwOutput{Boolean \Ids{isvalid}}
\KwRequire{$M_1, M_2$ are of same dimensions}
\BlankLine
\procedure(\procedurename{IsTransformable}{$M_{1},M_{2},t$}){
\State{Local $\Ids{isvalid} \gets \textrm{False}$}
\State{Local $\Ids{vars1} \gets \textrm{all variables from }M_1$}
\State{Local $\Ids{vars2} \gets \textrm{all variables from }M_2$}
\State{Local $\Ids{varssubs[1]} \gets \textrm{Variable replacement of }\Ids{vars1}\textrm{ to unique }a_i$}
\State{Local $\Ids{varssubs[2]} \gets \textrm{Variable replacement of }\Ids{vars2}\textrm{ to unique }b_i$}
\If{$\Call{Length}{vars1}<\Call{Length}{vars2}$}{
   \State{$\textrm{Return }\Ids{isvalid} \gets \textrm{False}$}}
\State{Local $\Ids{solset} \gets (M_1-M_2)\textrm{ after applying }\Ids{varsubs[1]}, \Ids{varsubs[2]}$}
\State{Local $\Ids{varset} \gets \{\textrm{all } a_{i}\}$}
\State{Local $\Ids{sols} \gets \Call{TimeConstrained}{\Call{Solve}{\Ids{solset}=\mathbf{0},\Ids{varset}},t,\{\}}$}
\ForEach{$s \in \Ids{sols}$}{
\If{$M_{1} = M_{2}\textrm{ after applying }\Ids{varsubs[1]}, \Ids{varsubs[2]} \textrm{ and } s$ }{\State{Local $\Ids{isvalid} \gets \textrm{True}$}
}}
\State{$\textrm{Return } \Ids{isvalid}$}
}
\end{algorithm}

\begin{algorithm}[H]
\small
\caption{Routine to transform R-Matrix closer to a triangular matrix \label{algr:Triformat}}
\KwDesc{Transforms the matrix towards an upper-triangular matrix structure upto R-Matrix symmetries}
\KwInput{Matrix $M_1$, Integer $N$}
\KwOutput{Matrix \Ids{M}}
\KwRequire{$M_1$ is a square matrix}
\BlankLine
\procedure(\procedurename{Triformat}{$M_{1},\;N$}){
\State{Local $\Ids{d} \gets \textrm{Dimension of }M_1$}
\State{Local $\Ids{rlist} \gets \Call{RMatrixInvariances}{M_1,N}$}
\State{Local $\Ids{weight} \gets \{ \Call{Triweight}{x}\;\forall\;x\in\Ids{rlist}\}$}
\State{Local $\Ids{weight} \gets \{x[1]+2^d*(\Call{Sum}{x[2]+x[3]})\;\forall\;x\in\Ids{weight}\}$}
\State{Local $\Ids{index} \gets \textrm{Position of highest value in }\Ids{weight}$}
\State{Local $\Ids{mout} \gets rlist[index]$}
\State{Return $\Ids{mout}$}
}
\end{algorithm}

\begin{algorithm}[H]
\small
\caption{Helper function to Triformat routine}
\KwDesc{Assigns a weight to a matrix to indicate the proximity with an upper triangular matrix}
\KwInput{Matrix $M_1$} 
\KwOutput{Number \Ids{n}}
\KwRequire{$M_1$ is a square matrix}
\BlankLine
\procedure(\procedurename{Triweight}{$M_{1}$}){
\State{Local $\Ids{d}\gets \textrm{Dimension of }M_{1}$}
\State{Local $\Ids{w1}\gets \frac{\textrm{Sum of upper triangular elements of }S(M_1)+1}{\textrm{Sum of lower triangular elements of }S(M_1)+1}$}
\BlankLine
\State{\textsc{mask}(s) =  $\left[(1+s)d+s+(-1)^{s}j-\mid i-j\mid\right]_{ij},\;\;1\leq i,j \leq d$}
\BlankLine
\State{Local $\Ids{w2}\gets \textrm{Table of sum of every upper diagonal terms of }\Call{Mask}{1}\circ S(M_1)$}
\State{Local $\Ids{w3}\gets \textrm{Table of sum of every lower diagonal terms of }\Call{Mask}{0}\circ S(M_1)$}
\State{Local $\Ids{n}\gets\{\Ids{w1},\Ids{w2},\Ids{w3}\}$}
\State{Return $\Ids{n}$}
}
\end{algorithm}

\begin{algorithm}[H]
\small
\caption{Routine to check if one R-Matrix is similar to another \label{algr:IsSimilar}}
\KwDesc{Checks if constant matrix $M_1$ is similar to $M_2$ upto variable substitution}
\KwInput{Matrix $M_1, M_2$, Integer $t$ (time, seconds)}
\KwOutput{Boolean \Ids{isvalid}}
\KwRequire{$M_1, M_2$ are of dimension $4\times 4$}
\BlankLine
\procedure(\procedurename{IsSimilar}{$M_{1},\;M_{2},\;t$}){
\State{Local $\Ids{isvalid} \gets \textrm{False}$}
\State{Local $Q = \left(\begin{array}{cc}q_1 & q_2 \\ q_3 & q_4\end{array}\right)\otimes\left(\begin{array}{cc}q_1 & q_2 \\ q_3 & q_4\end{array}\right)$}
\State{Local $\Ids{vars1} \gets \textrm{all variables from }M_1$}
\State{Local $\Ids{vars2} \gets \textrm{all variables from }M_2$}
\State{Local $\Ids{varssubs[1]} \gets \textrm{Variable replacement of }\Ids{vars1}\textrm{ to unique }a_i$}
\State{Local $\Ids{varssubs[2]} \gets \textrm{Variable replacement of }\Ids{vars2}\textrm{ to unique }b_i$}
\If{$\Call{Length}{vars1}+4<\Call{Length}{vars2}$}{
   \State{$\textrm{Return }\Ids{isvalid} \gets \textrm{False}$}
}
\State{Local $\Ids{solset} \gets (Q\cdot M_1\cdot Q^{-1}-M_2)\textrm{ after applying }\Ids{varsubs[1]}, \Ids{varsubs[2]}$}
\State{Local $\Ids{varset} \gets \{\textrm{all } a_{i} \textrm{ and } q_{1}, q_{2}, q_{3}, q_{4}\}$} 
\State{Local $\Ids{sols} \gets \Call{TimeConstrained}{\Call{Solve}{\Ids{solset}=\mathbf{0},\Ids{varset}},t,\{\}}$}

\ForEach{$s \in \Ids{sols}$}{
\If{$Q.M_{1}.Q^{-1} = M_{2}\textrm{ after applying }\Ids{varsubs[1]}, \Ids{varsubs[2]} \textrm{ and } s$ }{
   \State{Local $\Ids{isvalid} \gets \textrm{True}$}
}}
\State{$\textrm{Return } \Ids{isvalid}$}
}
\end{algorithm}